\documentclass[]{interact}

\usepackage{booktabs}                      
\usepackage{graphicx}                      
\usepackage{hyperref}                      
\usepackage{subcaption}                    
\usepackage{algorithm, algpseudocode, bm}  
\usepackage{soul}                          
\usepackage{mathtools}                     
\usepackage{tikz}                          
\usepackage{cleveref}                      
\usepackage{fontawesome}                   
\let\olddoi\doi
\renewcommand{\doi}[1]{{\color{blue}\olddoi{#1}}} 

\definecolor{lime}{HTML}{A6CE39}
\DeclareRobustCommand{\orcidicon}{
	\begin{tikzpicture}
	\draw[lime, fill=lime] (0,0) 
	circle [radius=0.16] 
	node[white] {{\fontfamily{qag}\selectfont \tiny ID}};
	\draw[white, fill=white] (-0.0625,0.095) 
	circle [radius=0.007];
	\end{tikzpicture}
	\hspace{-2mm}
}

\definecolor{linkcolor}{rgb}{0.1216,0.4667,0.7059}
\definecolor{twitterblue}{rgb}{0.1137,0.6314,0.949}
\newcommand{\codeicon}{{\color{linkcolor}\faFileCodeO}}  
\newcommand{\zoomicon}{{\color{linkcolor}\faSearch}}  

\newcommand{\nhigh}{n_h}
\newcommand{\nlow}{n_l}
\newcommand{\nmaxhigh}{\nhigh^\text{max}}
\newcommand{\nmaxlow}{\nlow^\text{max}}

\newcommand{\correlation}{r}
\newcommand{\costratio}{\phi}

\newcommand{\doehigh}{H}
\newcommand{\doelow}{L}

\newcommand{\norm}[1]{\left\lVert #1 \right\rVert}
\newcommand{\numreps}{I}

\DeclarePairedDelimiter\abs{\lvert}{\rvert}%
\DeclareMathOperator*{\argmin}{arg\,min}

\newcommand{\gradient}{\frac{\beta_{h}}{\beta_{l}}}
\newcommand{\dgradient}{\dfrac{\beta_{h}}{\beta_{l}}}
\newcommand{\nhighinit}{n_{h,0}}
\newcommand{\nlowinit}{n_{l,0}}

\newcommand{\m}[1]{{\ensuremath{\mathrm{#1}}}}

\newcommand{\sectionbreak}{\clearpage}

\UseRawInputEncoding

\graphicspath{{img/}}

\foreach \x in {A, ..., Z}{%
	\expandafter\xdef\csname orcid\x\endcsname{\noexpand\href{https://orcid.org/\csname orcidauthor\x\endcsname}{\noexpand\orcidicon}}
}

\newcommand{\codelink}[1]{\href{https://github.com/sjvrijn/multi-level-co-surrogates/blob/v1/scripts/processing/#1.py}{\codeicon}\,}
\newcommand{\zoomlink}[1]{\href{https://figshare.com/articles/figure/Finding_Efficient_Trade-offs_in_Multi-Fidelity_Response_Surface_Modeling_#1}{\zoomicon}\,}

\newcommand{\degree}{^{\circ}}

\begin{document}

\title{Finding Efficient Trade-offs in Multi-Fidelity Response Surface Modeling}











\author{
    \name{
        Sander van Rijn\textsuperscript{a}\orcidA\thanks{\faEnvelopeO\ Email: s.j.van.rijn@liacs.leidenuniv.nl}
        \and Sebastian Schmitt\textsuperscript{b}\orcidB
        \and Matthijs van Leeuwen\textsuperscript{a}\orcidC
        \and Thomas B\"ack\textsuperscript{a}\orcidD
    }
    \affil{
        \textsuperscript{a}LIACS, Leiden University, The Netherlands;
        \textsuperscript{b}Honda Research Institute Europe GmbH, Offenbach am Main, Germany
    }
}


\begin{abstract}
In optimization approaches to engineering applications, time-consuming simulations are often utilized which can be configured to deliver solutions for various fidelity (accuracy) levels. 
It is common practice to train hierarchical surrogate models on the objective functions in order to speed-up the optimization process.
These operate under the assumption that there is a correlation between the different fidelities 
that can be exploited to cheaply gain information.
However, limited guidelines are available to help divide the available computational budget between multiple fidelities in practice.
In this article we evaluate a range of different choices for a two-fidelity setup that provide helpful intuitions about this trade-off. 
We present a heuristic method based on subsampling from an initial Design of Experiments to find a suitable division of the computational budget between the fidelity levels.
This enables the setup of multi-fidelity optimizations which utilize the available computational budget efficiently, independent of the multi-fidelity model used.

\end{abstract}
\begin{keywords}
    multi-fidelity simulation problems;
    design of experiments;
    error grids;
    hierarchical surrogate models;
    co-kriging
\end{keywords}



\maketitle              


\section{Introduction}

    When dealing with simulation based optimization problems in engineering applications, the runtime cost of each evaluation is typically the most restrictive aspect of a successful approach. Surrogate models are often used to reduce the total computational load by learning trends from previous evaluations. But the computational cost for single evaluations have grown too high in many modern problems for such approaches to obtain enough information necessary to train an accurate model in reasonable time.
    
    Many such problems offer tunable accuracy and can therefore be classified as either arbitrarily tunable \emph{variable-fidelity}, or discretely tunable \emph{multi-fidelity} problems. In the following we focus on multi-fidelity problems specifically. Supplementing accurate high-fidelity information with cheaper low-fidelity information is regularly done by incorporating \emph{hierarchical} (co-)surrogate models based on work by \cite{kennedy_predicting_2000}, such as co-kriging~\cite{forrester_multi-fidelity_2007} and co-RBF~\cite{durantin_multifidelity_2017}. These have been successfully applied in, e.g., the design of ships~\cite{pellegrini_multi-fidelity_2016-1} airfoils~\cite{liu_sequential_2018}, satellites \cite{shi_multi-fidelity_2020}, additive manufacturing~\cite{zhou_accelerating_2019}, and fire start determination~\cite{li_multi-fidelity_2019}.

    However, it remains unclear under which conditions the inclusion of low-fidelity information in hierarchical surrogate models is actually beneficial. While previous research has shown that the correlation between high- and low-fidelity response surfaces should be fairly high (i.e., sample correlation coefficient $\correlation>0.9$~\cite{toal_considerations_2015,fernandez-godino_review_2016}), a high correlation by itself is no guarantee for achieving added benefit of multi-fidelity models. That is, regardless of correlation, individual response surface landscapes still have a substantial impact on the final accuracy of the trained models.

    Furthermore, even if a model is beneficial, how to best distribute the available computational budget between the fidelity levels is still an open question. Prior work has included experiments where models based on multiple sample sizes were compared, but most present only a limited selection of combinations, as can be seen in, e.g., the overviews by \cite{fernandez-godino_review_2016,fernandez-godino_linear_2019}.
    Common heuristics for deciding on this division rely either on the cost ratio between fidelities or otherwise use expected information gains~\cite{guo_design_2020,moss_mumbo_2020,huang_sequential_2006,ryou_multi-fidelity_2020,belakaria_multi-fidelity_2020}.
    Of these, the former do not use any function information, while the latter are designed to be used in iterative optimization, not to gain general understanding.

    In this work we empirically explore how to distribute additional computational budget over two fidelities.
    We focus on one-shot Design of Experiments (DoEs), by enumerating all possible combinations of a wide range of low and high fidelity samples, and fit a hierarchical model and measure its accuracy for each setup.
    Our approach is similar to a study by \cite{durantin_multifidelity_2017} but with a much finer granularity, more sample combinations, and many more benchmark functions.
    By analyzing model accuracy as a function of the DoE sizes and for various benchmark functions, we aim to provide general insight into the behavior of this trade-off.

    We recognize that using an enumeration procedure to obtain this information is far too computationally expensive in terms of problem evaluations for practical settings.
    Therefore, we present a method using subsampling that draws smaller DoEs from an initial DoE to avoid performing any new evaluations.
    Using these subsampled DoEs we approximate the accuracy trend results and show a high correlation between these results and those from the original full enumeration.
    
    We present a heuristic which utilizes the information gained from the subsampling approach to predict a beneficial split of the number of high and low-fidelity samples for a given computational budget.
    This allows for an efficient use of the available computational resources for the multi-fidelity modeling approach to optimization.

    All files for this work are archived on Zenodo~\cite{mlcs_code_zenodo,mlcs_data_zenodo}. The \codeicon\, and \zoomicon\, icons under each figure link to the source code on GitHub~\cite{van_rijn_multi-level-co-surrogates_2020} and full versions on FigShare \cite{figshare_figures} respectively.


\section{Background}
\label{sec:background}

    In this section we define some terms and methods that are used in the remainder of this article.



    \subsection{Multi-Fidelity Problems}
    \label{subsec:bg-mf-problem}

        A multi-fidelity problem is an optimization/simulation problem that is available in multiple \emph{fidelities}, i.e., accuracy levels. In real-world Computational Fluid Dynamics (CFD) simulations of, e.g., airfoils, these fidelities could correspond to different mesh sizes or simulation types. A \emph{low}-fidelity simulation would use a coarse mesh or potential flow solver, and thus give lower accuracy, but be faster to calculate, while a \emph{high}-fidelity simulation would use a finer mesh or Reynolds-averaged Navier-Stokes (RANS) simulation and therefore be more accurate while taking longer to calculate.

        In the following, we will use $f_h: {\cal X} \rightarrow {\cal Y}$ and $f_l: {\cal X} \rightarrow {\cal Y}$ to denote the high- and low-fidelity levels of a simulator, abstractly represented by the function $f$ that maps input vectors $\mathbf{x} \in {\cal X}$ onto outputs $\mathbf{y} \in {\cal Y}$.

    \subsection{Additive Hierarchical Surrogate Models}
    \label{subsec:bg-hierarchical}

        To make use of the multiple sources of information in a multi-fidelity problem, an additive model-structure has been proposed by \cite{kennedy_predicting_2000}:
        \begin{equation}
            z_h(\mathbf{x}) = \rho z_l(\mathbf{x}) + \delta(\mathbf{x})
        \end{equation}
        Here, $z_h(\mathbf{x})$, and $z_l(\mathbf{x})$ are the high- and low-fidelity surrogate models at point $\mathbf{x}$ respectively, for approximating $f_h(\mathbf{x})$ and $f_l(\mathbf{x})$. $\rho$ is a regression parameter, and $\delta(\mathbf{x})$ is the difference model at point $\mathbf{x}$, which improves the low-fidelity prediction by additive correction.

        Without loss of generality, we consider simplified additive models where the regression parameter $\rho = 1$, to limit algorithmic complexity. While this may reduce the achievable accuracy of the models, it is of no relevance to the concepts we introduce. Learning this parameter will most likely only further increase the performance of our approach. We create independent models for $z_l$ and $\delta$, where $z_l$ models the lowest accuracy information source  $f_l$, and a separate model $\delta$  predicts the differences between the high- and low-fidelity responses $f_h(\mathbf{x})-f_l(\mathbf{x})$. Specifically, we will be using Kriging models using the Mat\'ern kernel in this article, although the proposed method does not depend on this particular choice and could use other models.


    \subsection{Multi-Fidelity Design of Experiments}
    \label{subsec:bg-mf-DoE}


        A standard approach for systematically sampling a set of input parameter configurations in order to create a dataset of input-output pairs of given function is referred to as Design of Experiments (DoE) ~\cite{douglas_c_montgomery_design_2019}.
        The goal when choosing such a dataset is to cover the input-space of the function in such a way that the created model is as good as it can be, whether on a local or global scale.
        How large the search space is and how much computational effort can be expended on this depends on each individual problem setting. However, a full factorial design (i.e., grid search) is usually out of the question due to relatively high dimensionality and high computational cost.
        In this work, we use the common Latin Hypercube Sample (LHS) strategy. This technique tries to create a sample such that the samples are evenly distributed over the search space, while avoiding the reuse of coordinate values for a dimension.

        In a multi-fidelity setting, where we train a difference model between high and low fidelity functions, we would like to have overlapping DoEs for low and high fidelity models where all high fidelity samples are also included in the low fidelity DoE. Additionally, each DoE should still cover the search space efficiently and therefore should be an approximate LHS itself.

\begin{figure}
            \centering
            \begin{subfigure}[t]{.33\textwidth}
                \centering
                \includegraphics[width=\textwidth]{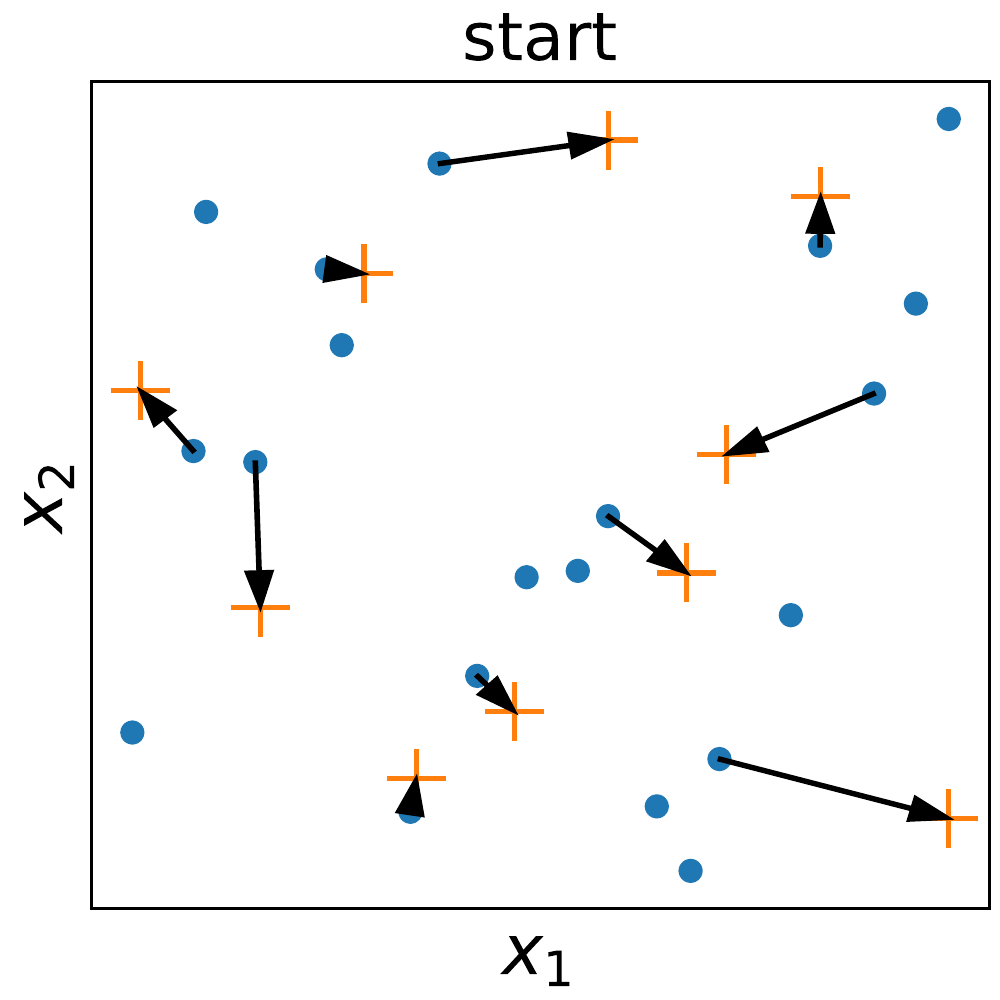}
            \end{subfigure}
            \begin{subfigure}[t]{.33\textwidth}
                \centering
                \includegraphics[width=\textwidth]{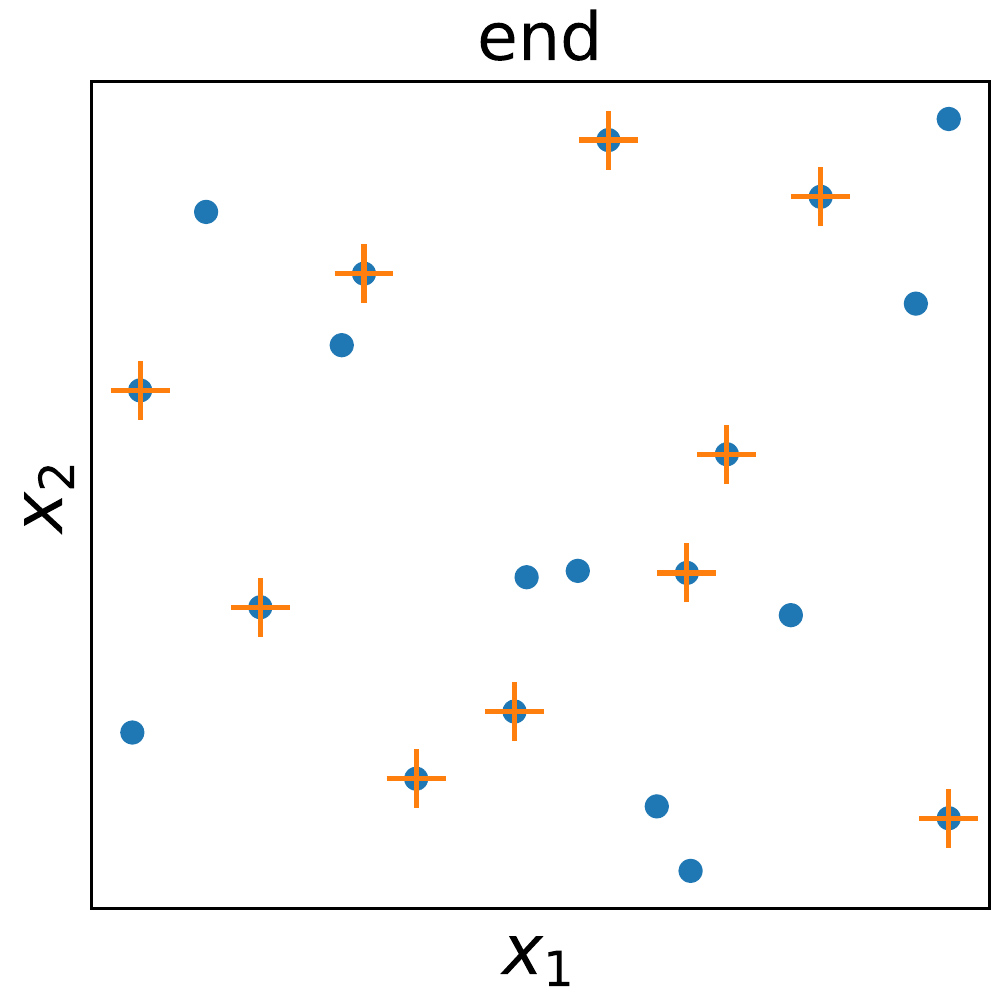}
            \end{subfigure}
            \caption{
                \textbf{Illustration of \Cref{alg:mf-doe}}.
                A 2-dimensional DoE with $\nhigh = \abs{\doehigh} = 10$ and $\nlow = \abs{\doelow} = 20$
                \codelink{2020-07-29-illustrated-bi-fid-doe}
                \zoomlink{2020-07-29-illustrated-bi-fid-doe/14060912}
            }
            \label{fig:bi-fid-doe-sampling}
        \end{figure}

        To achieve this, we use the procedure from \cite{le_gratiet_multi-fidelity_2013} to generate DoEs for the hierarchical models, as outlined in \Cref{alg:mf-doe}. In lines 1-2, two separate LHSs are generated. Then, for each high-fidelity point, the closest low-fidelity point is replaced by that high-fidelity point (while-loop in lines 4-9). Given the desired sizes, this method returns both a high-fidelity LHS $\doehigh$, and a more spread-out low-fidelity sample $\doelow$ which is still roughly an LHS itself. As the final outcome we arrive at a DoE  which is the union of the sets $\doelow$ and $\doehigh \subset \doelow$ of low and high fidelity samples, respectively, which we denote as DoE$(\doehigh, \doelow)$ and where the exact differences $f_h(\mathbf{x}) - f_l(\mathbf{x})$  can be computed for all $\mathbf{x} \in \doehigh$ and used as training set for the difference model $\delta$. \Cref{fig:bi-fid-doe-sampling} illustrates this procedure.
        
        \begin{algorithm}[!ht]
            \begin{algorithmic}[1]
                \Require $\nlow \geq \nhigh + 1$
            
                \State{$\doehigh{} \leftarrow$ LHS$(\nhigh)$} \Comment{Independent samples per fidelity}
                \State{$\doelow{} \leftarrow$ LHS$(\nlow)$}
                \State{$\doelow{}^\prime, \doehigh{}^\prime \leftarrow \emptyset$}
            
                \While{$\doehigh{}$ not empty}
                    \State{$\mathbf{h}, \mathbf{l} \leftarrow \argmin_{\mathbf{h} \in \doehigh{}, \mathbf{l} \in \doelow{}} \norm{\mathbf{h} - \mathbf{l}} $} \Comment{Find closest pair}
                    \State{$\doehigh{}^\prime \leftarrow \doehigh{}^\prime \bigcup \mathbf{h}$}
                    \State{$\doelow{}^\prime \leftarrow \doelow{}^\prime \bigcup \mathbf{h}$} \Comment{Effectively adjust $\mathbf{l}$ to $\mathbf{h}$}
                    \State{Remove $\mathbf{h}, \mathbf{l}$ from $\doehigh{}, \doelow{}$}
                \EndWhile
                \State{$\doelow{}^\prime \leftarrow \doelow{}^\prime \bigcup \doelow{}$}\Comment{Add remaining low-fid points}
                \State\Return{$\doehigh{}^\prime, \doelow{}^\prime$}
            \end{algorithmic}
            \caption{Multi-Fidelity LHS \cite{le_gratiet_multi-fidelity_2013}}
            \label{alg:mf-doe}
        \end{algorithm}  
 
\section{Problem Statement}
\label{sec:problem}

    The fundamental question we address in this work is: how should a given additional computational budget be distributed among multiple fidelities? Especially, in the context of computationally expensive simulation problems where the overall evaluation budget is constrained, this is a highly relevant question. As a starting point we choose to address the question of how to split the high and low fidelity samples for the DoE$(\doehigh, \doelow)$.

    The answer to this question depends on which fidelity provides the most information for its computational cost. An additional high-fidelity sample in a so-far unexplored area will definitely improve the model's accuracy. But if some number of low-fidelity samples can improve the model more with equal or lower computational cost, that might be a better choice. How much information is gained by adding another sample for a specific fidelity depends heavily on the number of samples of that fidelity already present, and on the chosen model's capacity to capture the problem's response surface.

    An important quantity in determining the split between the number of high- and low-fidelity evaluations is given by the  \emph{cost ratio} $\costratio = c_l/c_h \in (0, 1)$, where $c_h$ and $c_l$ are typical computation times associated with high- and low-fidelity evaluations, respectively.
    The problem can thus be stated as follows: given a fixed evaluation budget $b$ (which is measured in high-fidelity evaluation times) and cost ratio $\costratio$, what are the optimal numbers of high- and low-fidelity evaluations, i.e.\ the optimal division ratio $\frac{\nhigh}{\nlow}$ that minimizes the model error and which respects the constraint that the budget is not exceeded, i.e.\ $\nhigh+\costratio \nlow\leq b$.
    As the model error, we take the Mean Squared Error (MSE) of response surface model $z$ compared to the true function value of the highest fidelity level  $f_h$, evaluated on a given test set $x\in T$:
    \begin{equation}
        \label{eq:mse}
        MSE(z, T) = \sum_{\mathbf{x} \in T} \dfrac{( z(\mathbf{x}) - f_h(\mathbf{x}))^2}{\abs{T}}.
    \end{equation}

    The model error is expected to have a nontrivial and in general non-linear behavior as a function of the division ratio.
    For the case consisting only of high fidelity evaluations, the low number of overall samples probably leads to a rather large error. On the other extreme, using only low-fidelity evaluations will also not produce an accurate model as no actual information about the true function is used. In the intermediate region,  with some low-fidelity samples at the expense of a few high-fidelity evaluations it is expected to obtain a reduced model error. 
    The details of this trade-off strongly depend on various aspects of the problem, like the structure of the high-fidelity function and the similarity between the low- and the high-fidelity functions. However, in order to learn the dominant behavior of the error as function of the number of high- and low-fidelity samples we will later employ a simple fit to the error. This will enable us to extract the global trend and allows us to formulate a heuristic which guides the distribution of additional computational budget.


\section{Method}
\label{sec:method}

    \subsection{Enumeration of Multi-Fidelity DoE Sizes}
    \label{subsec:method-brute-force-enumeration}

        \begin{figure*}
            \centering
            \begin{subfigure}[t]{.49\textwidth} 
                \centering
                \includegraphics[width=\textwidth]{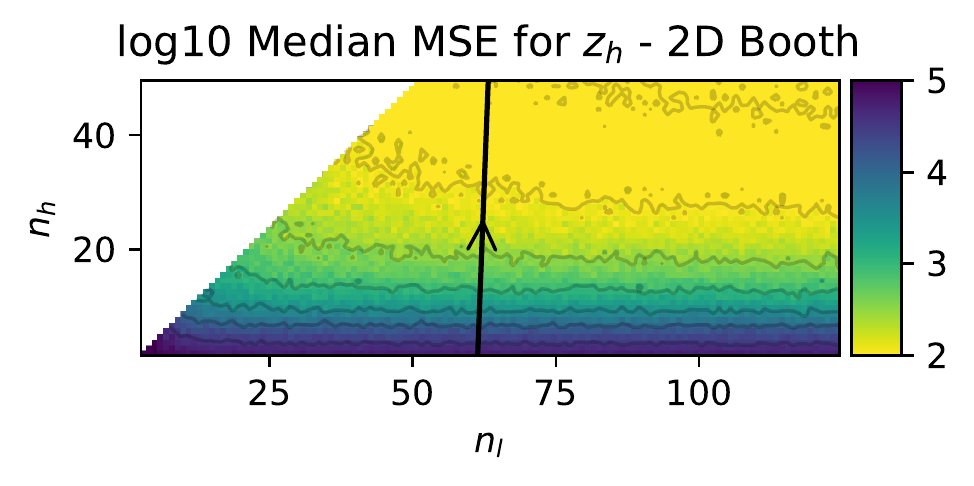}
                \caption{}
                \label{fig:example-heatmaps:booth}
            \end{subfigure}
            \begin{subfigure}[t]{.49\textwidth}
                \centering
                \includegraphics[width=\textwidth]{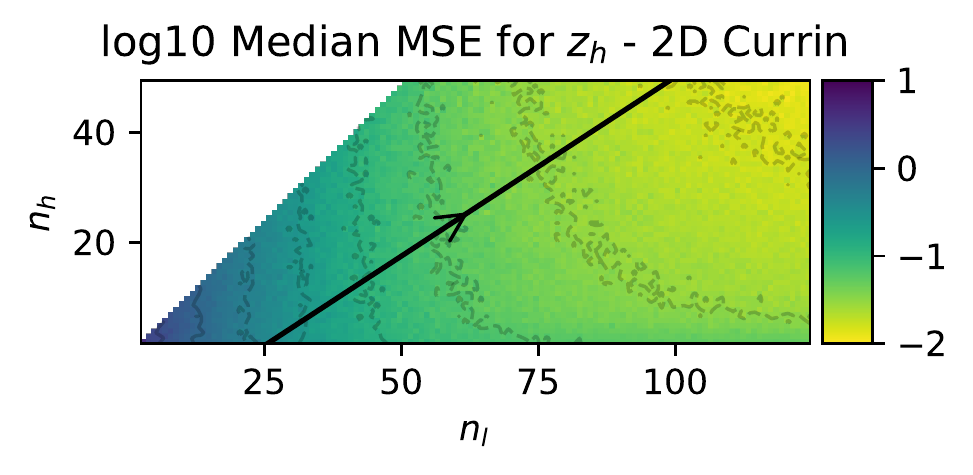}
                \caption{}
                \label{fig:example-heatmaps:currin}
            \end{subfigure}
            \begin{subfigure}[t]{.49\textwidth}
                \centering
                \includegraphics[width=\textwidth]{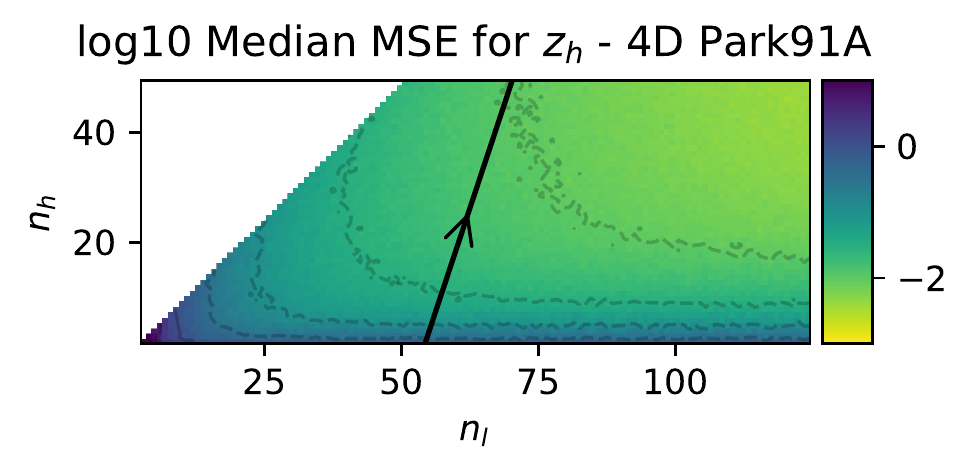}
                \caption{}
                \label{fig:example-heatmaps:park91a}
            \end{subfigure}
            \begin{subfigure}[t]{.49\textwidth}
                \centering
                \includegraphics[width=\textwidth]{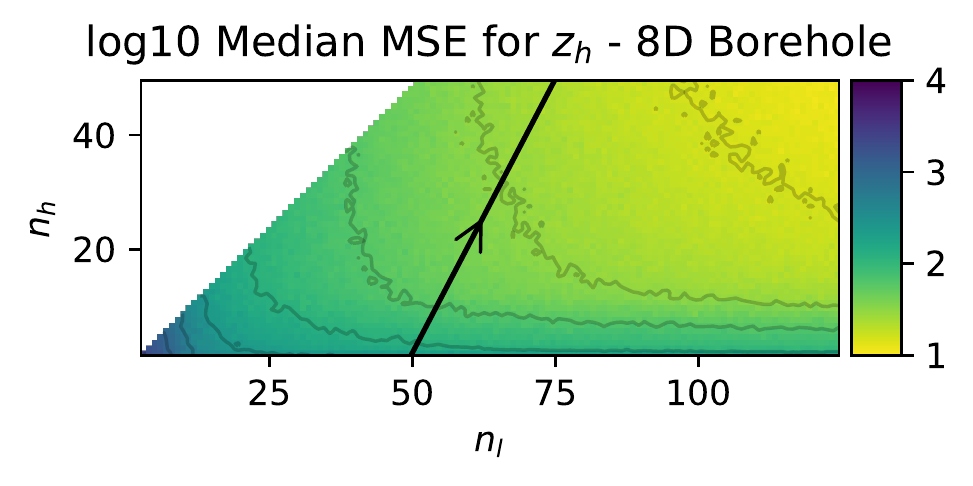}
                \caption{}
                \label{fig:example-heatmaps:borehole}
            \end{subfigure}
            \caption{
                \textbf{Error Grids} Heatmaps of the $\log10$ of the hierarchical model MSE for varying DoE sizes, shown as the median over $\numreps = 50$ iterations, on four benchmark problems. The black arrow shows the global gradient direction as described in \Cref{subsec:method-angle-quantification}. Overall, it is clear that adding more samples improves model accuracy, either additional high-fidelity samples (vertical) or low-fidelity samples (horizontal)
                \codelink{2019-09-19-plot-error-grids}
                \zoomlink{2019-09-mse-nc/14060957}
            }
            \label{fig:example-heatmaps}
        \end{figure*}  

        To fully examine the trade-off between the number of high- and low-fidelity samples, model accuracy information needs to be obtained for many possible combinations. We gather such information by empirically performing a full enumeration of all possible combinations $(\nlow,\nhigh)$ for $2<\nhigh<\nmaxhigh$ and $ \nhigh+1<\nlow< \nmaxlow$. For each pair $(\nlow,\nhigh)$ we train multiple ($I=50$) hierarchical multi-fidelity models to collect  some statistics and evaluate the errors on an independent test set $T$. We refer to the tables of errors for the complete enumeration DoEs as \emph{error grids}.

        \begin{algorithm}[!ht]
            \begin{algorithmic}[1]
                \Require $N$-dimensional multi-fidelity problem ($f_h, f_l$)
                \Require $\nmaxhigh, \nmaxlow$ \Comment{Maximum DoE size}
                \Require \numreps \Comment{Number of iterations}
                
                \State{$E \leftarrow \emptyset$} \Comment{Error Grid Storage}
                \State{$T \leftarrow$ LHS$(500 \cdot N)$} \Comment{Independent test set}
                
                \For{$\nhigh = 2 \dots \nmaxhigh$}
                    \For{$\nlow = (\nhigh+1) \dots \nmaxlow$}
                        \For{$i = 1 \dots \numreps$}
                            \State{$\doehigh{}, \doelow{} \leftarrow \text{MF-LHS}(\nhigh, \nlow)$} \Comment{\Cref{alg:mf-doe}}
                            \State{$Y_h, Y_l \leftarrow f_h(\doehigh), f_l(\doelow)$} \Comment{Evaluate}
                            \State{Train $z_h$ using $\doehigh, \doelow, Y_h, Y_l$}
                            \State{$E[\nhigh,\nlow,i] \leftarrow$ MSE($z_h, T$)} \Comment{\Cref{eq:mse}}
                        \EndFor
                    \EndFor
                \EndFor
                \State\Return{$E$}
            \end{algorithmic}
            \caption{Full Enumeration Error Grid}
            \label{alg:brute-force}
        \end{algorithm}  

        \Cref{alg:brute-force} lists a pseudo-code representation of the procedure by which we obtain the error grids. 
        The size of the test set $T$ is set to $|T|=500 \cdot N$, where $N$ is the dimensionality of the search space (line 2). For each combination ($\nhigh, \nlow$), $I=50$ independent DoEs are sampled and the errors for the multi-fidelity model based on each DoE are evaluated and stored (lines 5--10).  
    
        The resulting error between the surrogate model and the true high fidelity function can be visualized in heatmaps of the error grids as shown in \Cref{fig:example-heatmaps}. We show the median error over the $I$ independent realizations of the DoEs as 2D heatmaps and as function of $\doehigh$ and $\doelow$. Our experiments showed that the distribution of the MSEs is exponential, so we use the $\log_{10}(\text{MSE})$ to better account for the different error scales. These error grids serve as the basis for our analysis as we can:
        \begin{itemize}
            \item examine the dependence of the model error as function of the division ratio between the number of high- and low-fidelity evaluations;
            \item examine how this dependency varies between multi-fidelity  problems; and
            \item identify the optimal division ratio for a given budget and problem.
        \end{itemize}

        However, the number of DoEs which need to be evaluated in this full enumeration procedure  is $N_\mathrm{DoE}=\numreps\, \nmaxhigh\, (\nmaxlow - \nmaxhigh / 2)$, which for $\numreps = 50$, $\nmaxhigh = 50$, $\nmaxlow = 125$ is a total of $N_\mathrm{DoE}=250\,000$ DoEs to sample and models to train. The number of high-fidelity function evaluations is consequently in the millions, which might be feasible for trivially computable benchmark problems, but will be prohibitively infeasible for real-world problems with higher computational demand.

    \subsection{Cross-Validated Subsampling of DoE Sizes}
    \label{subsec:method-subsampling}

        In this section we describe how to approximate error grids using only one fixed initial multi-fidelity DoE.  Given  one DoE ($\doehigh, \doelow$) with $\nhigh=|\doehigh|$ and  $ \nlow=|\doelow|$ samples, we can create a full error grid by reusing these evaluated samples and creating a set of smaller subsampled DoEs. Concretely, we subsample $\doehigh^\prime \subset \doehigh, \doelow^\prime \subset \doelow$ of size $(\nhigh^\prime, \nlow^\prime)$ such that the set of high fidelity samples $\doehigh^\prime$ is still a true subset of the set of low fidelity samples $\doelow^\prime$, i.e.,\ $\doehigh^\prime \subset \doelow^\prime$. For this, we first draw $\doehigh^\prime$ uniformly at random without replacement from the available $\doehigh$. Then, we take those samples as a start for $\doelow^\prime$, and add randomly chosen low-fidelity points until the desired size is reached (see \Cref{alg:subsample-doe}).

        \begin{algorithm}[!ht]
            \begin{algorithmic}[1]
                \Require Initial multi-fidelity DoE ($\doehigh$, $\doelow$)
                \Require Desired DoE size $\nhigh^\prime, \nlow^\prime$
                
                \State{$\doehigh^\prime \leftarrow$ uniform randomly choose $\nhigh^\prime$  samples from $\doehigh$}
                \State{$\doelow^\prime \leftarrow$ uniform randomly choose remaining $\nlow^\prime - \nhigh^\prime$ samples from $(\doelow \setminus \doehigh^\prime)$}
                \State{$\doelow^\prime \leftarrow \doelow^\prime \bigcup \doehigh^\prime$}
                \State\Return{$\doehigh^\prime, \doelow^\prime$}
            \end{algorithmic}
            \caption{Subsampling MF-DoE}
            \label{alg:subsample-doe}
        \end{algorithm}  

        Since the chosen high-fidelity DoE $\doehigh^\prime$ is a strict subset of the original DoE $\doehigh$, we can use some samples left out of the subsampled DoE $\doehigh^\prime$ to serve as test set $\doehigh^\text{test} = \doehigh \setminus \doehigh^\prime$ and calculate the error of the surrogate models for each DoE similar to cross-validation. The complete subsampling approach is summarized in pseudocode shown in \Cref{alg:subsampling}.

        \begin{algorithm}[!ht]
            \begin{algorithmic}[1]
                \Require $N$-dimensional multi-fidelity problem ($f_h, f_l$)
                \Require $\numreps$ \Comment{Number of iterations}
                 
                \State{$E \leftarrow \emptyset$} \Comment{Error Grid Storage}
                \State{$\doehigh, \doelow \leftarrow \text{MF-LHS}(\nmaxhigh, \nmaxlow)$} \Comment{\Cref{alg:mf-doe}}
                \State{$Y_h, Y_l \leftarrow f_h(\doehigh), f_l(\doelow)$} \Comment{Evaluate once}
                 
                \For{$\nhigh = 2 \dots \nmaxhigh - 1$}
                    \For{$\nlow = (\nhigh+1) \dots \nmaxlow$}
                        \For{$i = 1 \dots \numreps$}
                            \State{$\doehigh^\prime, \doelow^\prime \leftarrow$ Subsample $\doehigh,\doelow$} \Comment{\Cref{alg:subsample-doe}}
                            \State{$Y_h^\prime, Y_l^\prime \leftarrow$ values from $Y_h, Y_l$ for $\doehigh^\prime, \doelow^\prime$}
                            \State{Train $z_h$ using $\doehigh^\prime, \doelow^\prime, Y_h^\prime, Y_l^\prime$}
                            \State{$E[\nhigh,\nlow,i] \leftarrow$ MSE($z_h, \doehigh^\text{tst}$)} \Comment{$\doehigh^\text{tst}: \doehigh \setminus \doehigh^\prime$}
                        \EndFor
                    \EndFor
                \EndFor
                \State\Return{$E$}
            \end{algorithmic}
            \caption{Subsampling Error Grid Procedure}
            \label{alg:subsampling}
        \end{algorithm}  

    A comparison of the subsampling and full enumeration procedure is done in \Cref{app:subsampling}.

    \subsection{Angle of Gradient Quantification}
    \label{subsec:method-angle-quantification}

        The error grids provide intuitive information about the trade-off between the numbers of high- and low-fidelity samples. The contour lines give a very clear visual guidance in which direction of the $(\nlow,\nhigh)$-plane the accuracy of the surrogate models increases. To evaluate this behavior quantitatively, we use the gradient of the error with respect to the number of samples. If this gradient direction is predominantly along the $\nhigh$ direction, i.e.,\ it has an angle close to 90$\degree$ as measured from the horizontal $\nlow$ axis (see example in \Cref{fig:example-heatmaps}), this indicates that improvements in model quality mostly depend on additional high-fidelity information. However, if the error gradient angle is more horizontal, the benefit of adding low-fidelity information is larger. It is important to note that even when the angle is mostly vertical, e.g., 80$\degree$, adding low-fidelity information can still be beneficial as long as it is computationally much cheaper.

        In the following we use the direction of the error gradient to estimate the best split between high and low fidelity samples in order to reduce the modeling error. Even though the gradient is not consistent throughout the error grid, as can be seen by the curved contour lines, we can extract the global behavior by fitting a hyperplane through the $\log_{10}$ of the MSE data according to
        \begin{equation}
            \label{eq:lin-reg}
            \log_{10}(MSE) = \alpha + \beta_h \nhigh + \beta_l \nlow.
        \end{equation}
        Although clearly an approximation, the global direction provided by a simple linear model is sufficient for our purpose. From the linear fit, the global direction of the gradient direction of reducing error can be summarized intuitively by an angle
        \begin{equation}
            \label{eq:angle}
            \theta = \arctan\left(\frac{\beta_h}{\beta_l}\right).
        \end{equation}
        For the error grids in \Cref{fig:example-heatmaps}, for example, this results in angles of $\theta_{\m{Booth (2D)}}\approx88\degree$, $\theta_{\m{Currin (2D)}}\approx34\degree$, $\theta_{\m{Park91A (4D)}}\approx72\degree$ and $\theta_{\m{Borehole(8D)}}\approx63\degree$, respectively.
        We show confidence intervals of the calculated gradient angles in \Cref{fig:angle_vs_corr}, as determined using the calculations shown in \Cref{app:angle-ci}.


\section{Experiments}
\label{sec:experiments}

    \subsection{Replication of results and implementation details}
    \label{subsec:experiments-implementation}

        All source code of this work is available on GitHub~\cite{van_rijn_multi-level-co-surrogates_2020}, and archived on Zenodo~\cite{mlcs_code_zenodo,mlcs_data_zenodo}\, together with data files. The analyses are written in Python 3.6+, most notably using the packages \texttt{matplotlib}~\cite{hunter_matplotlib_2007}, \texttt{numpy}~\cite{van_der_walt_numpy_2011}, \texttt{scikit-learn}~\cite{pedregosa_scikit-learn_2011}, and \texttt{xarray}~\cite{hoyer_xarray_2017}, \href{https://github.com/sjvrijn/multi-level-co-surrogates/blob/v1/requirements.txt}{\codeicon}\,. Reproducibility is guaranteed by using a single fixed random seed for globally used random values such as the test set $T$ and the initial DoE used for subsampling \href{https://github.com/sjvrijn/multi-level-co-surrogates/blob/v1/scripts/experiments/experiments.py\#L35}{\codeicon}\,. 
        All experiments use $\numreps = 50$ iterations.

    \subsection{Benchmark functions}
    \label{subsec:experiments-benchmarks}
        
        \begin{figure}
            \centering
            \begin{subfigure}[t]{\textwidth}
                \centering
                \includegraphics[width=.32\textwidth]{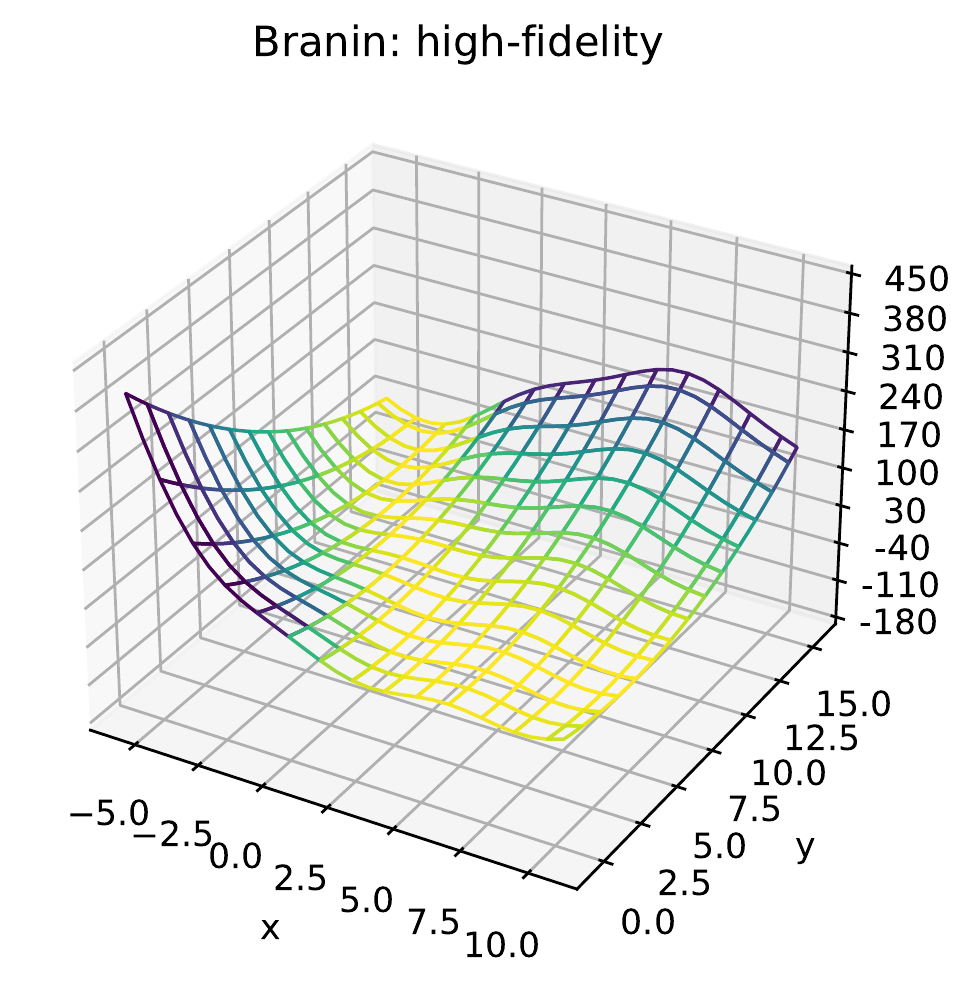}
            \end{subfigure}
            \begin{subfigure}[t]{.32\textwidth}
                \centering
                \includegraphics[width=\textwidth]{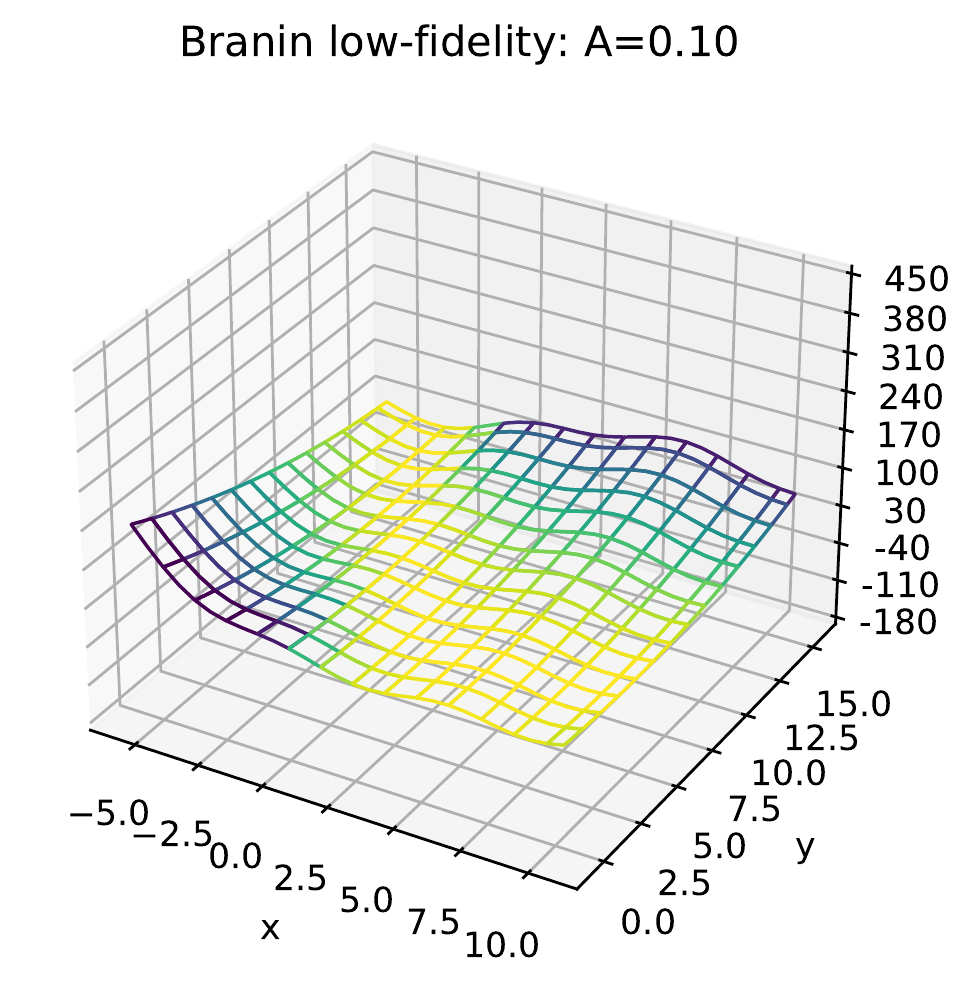}
            \end{subfigure}
            \begin{subfigure}[t]{.32\textwidth}
                \centering
                \includegraphics[width=\textwidth]{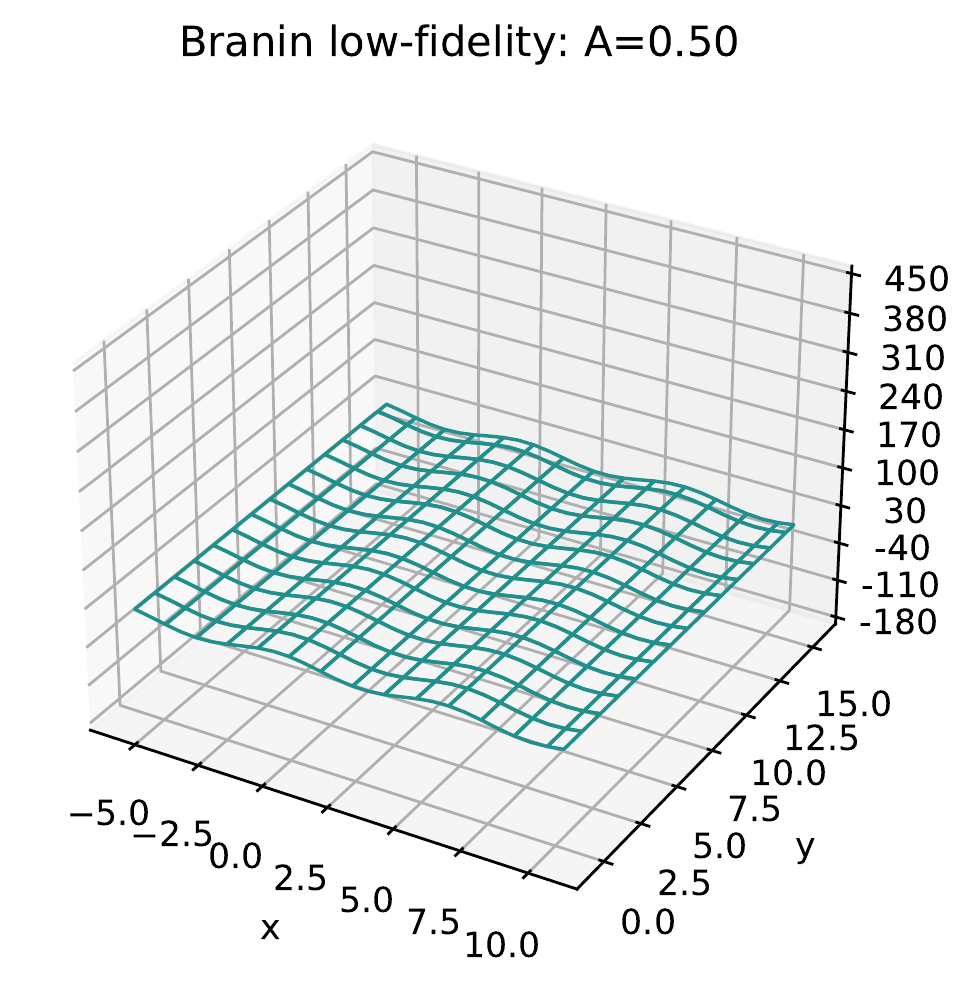}
            \end{subfigure}
            \begin{subfigure}[t]{.32\textwidth}
                \centering
                \includegraphics[width=\textwidth]{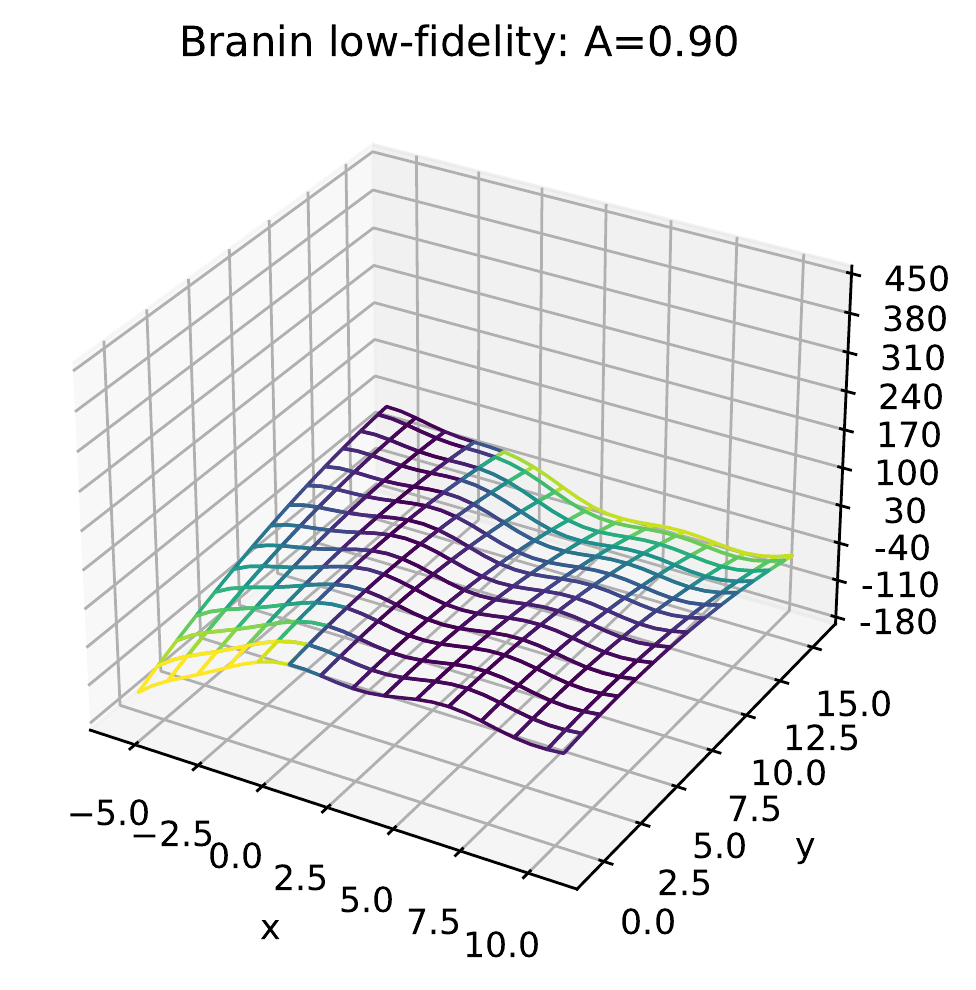}
            \end{subfigure}
            \caption{\textbf{Adjustable multi-fidelity function example} showing the adjustable Branin function. Top: high-fidelity. Bottom: low-fidelity for $A=0.1, 0.5, 0.9$ respectively}
            \label{fig:adjustable_example}
        \end{figure}  
        
        \begin{figure}
            \centering
            \includegraphics[width=.66\textwidth]{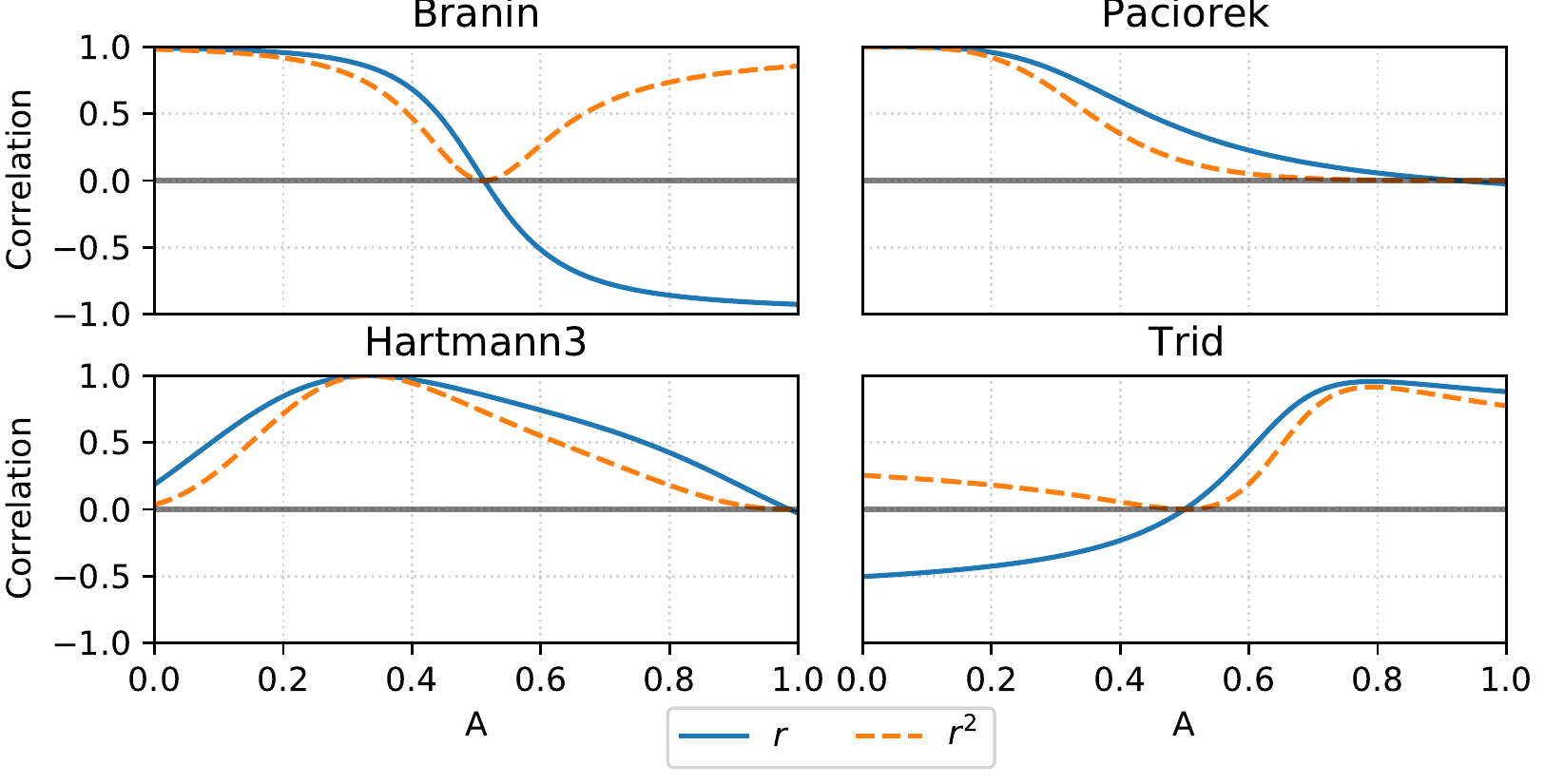}
            \caption{
                Correlation between high- and low-fidelity for adjustable 2D Branin, 2D Paciorek, 3D Hartmann3 and 10D Trid functions as function of parameter $A$
                \codelink{2019-10-30-correlation-table}
                \zoomlink{2019-10-correlation-exploration/14061014}
            }
            \label{fig:adjustable_correlations}
        \end{figure}  


        The used benchmark functions from the \texttt{mf2}~\cite{van_rijn_mf2_2020} package range from one dimensional (1D), such as the Forrester function, to 10D such as the Trid function, with a majority of 2D functions such as Bohachevsky, Currin and Six-Hump Camelback. This collection contains several different problem landscapes and all are previously used in the literature. With the exception of the 2D Branin function, the correlation between between their high- and low-fidelity functions are all above 0.7.

        In particular, we focus on the adjustable benchmark functions previously proposed in Section 3 of~\cite{toal_considerations_2015}: the 2D \emph{adjustable} Branin\footnote{Since Toal's adjustable Branin function differs from the non-adjustable version by \cite{dong_multi-fidelity_2015}, we explicitly differentiate between them by referring to Toal's version as \emph{adjustable}}, 2D Paciorek, 3D Hartmann3, and the 10D Trid function. The low-fidelity functions of these benchmarks include a tuning parameter $A$, which controls the correlation between high- and low-fidelity for these functions. \Cref{fig:adjustable_example} shows an example of this, although the exact influence of $A$ depends on the specific function, and the relationship between $A$ and the correlation for these functions is shown in \Cref{fig:adjustable_correlations}. For all functions, correlation can be tuned to any value between maximally positive ($\correlation \approx 1$) and absent ($\correlation \approx 0$). Additionally, for the Branin and Trid functions, this range extends to negative correlations ($\correlation \approx -1$). The explicit functional forms can be found in the article by \cite{toal_considerations_2015}.

    \subsection{Error Gradient Angle Analysis}
    \label{subsec:experiments-gradient-angle-analysis}

\begin{figure}
            \centering
            \includegraphics[width=.66\textwidth]{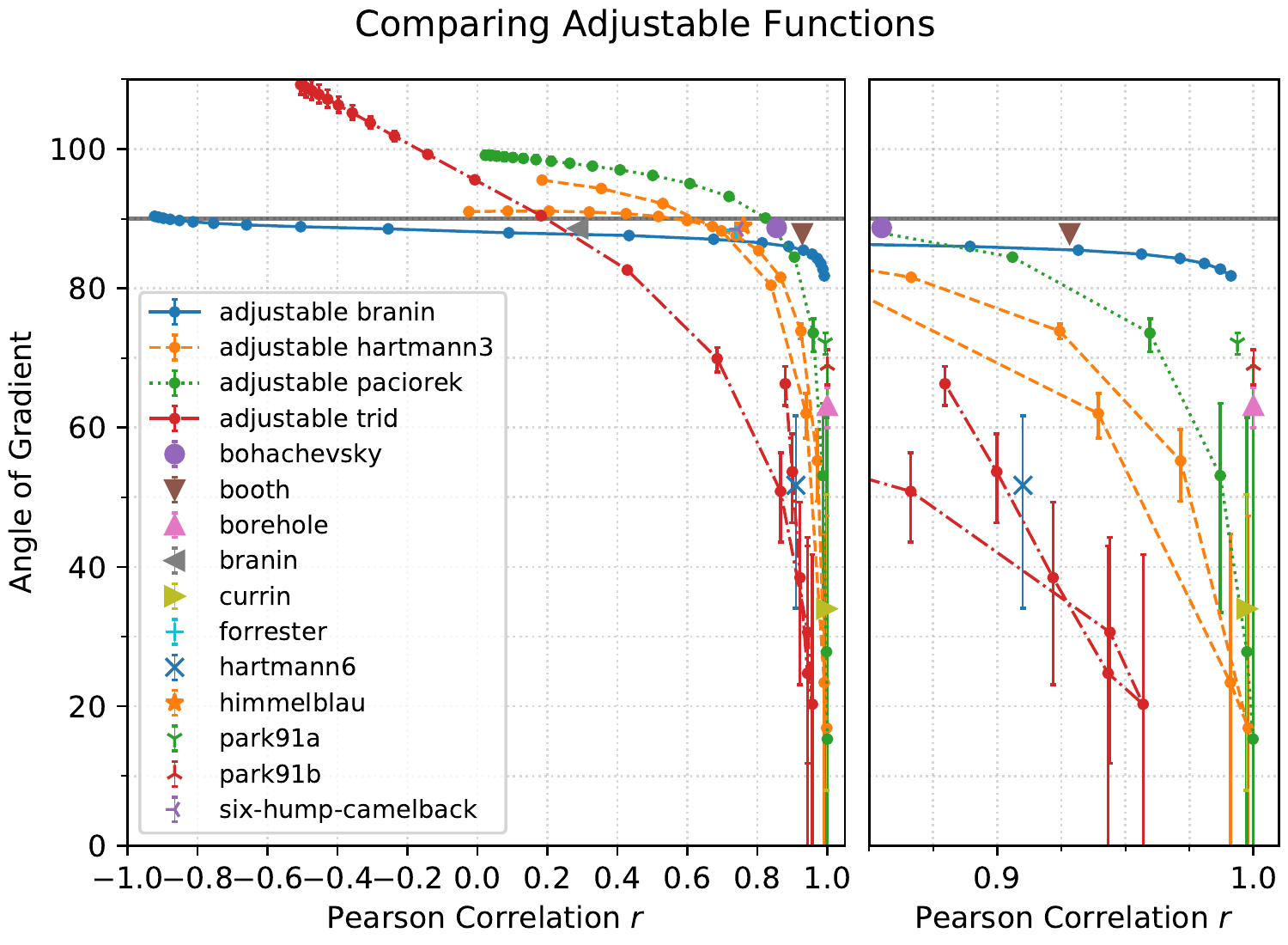}%
            \caption{
                \textbf{Angle as function of correlation} illustrated for all functions in the \texttt{mf2}~\cite{van_rijn_mf2_2020} package. The left side shows the complete range $-1 \leq r \leq 1$, while the right side highlights the highly correlated region $0.85 \leq r \leq 1$. Single markers are used for non-adjustable functions such as Booth and Borehole, while the lines with markers show various parameter values for the adjustable functions where the line connects points with adjacent values of $A$. Error bars show the CI as defined in \Cref{eq:angle-ci}
                \codelink{2020-02-19-adjustable-gradients}
                \zoomlink{2020-02-19-adjustable-gradients/14061017}
            }
            \label{fig:angle_vs_corr}
        \end{figure}

        The full enumeration procedure described in \Cref{subsec:method-brute-force-enumeration} was run for all \texttt{mf2} functions, using parameter values $A \in [0, 0.05, \ldots, 0.95, 1.0]$\footnote{For $A=0$, the high- and low-fidelity versions of the Paciorek function are identical, so we omit it.} for the adjustable functions. For each function, the gradient direction and error gradient angle was estimated using the linear fit procedure described in \Cref{subsec:method-angle-quantification}. The resulting angles, along with the confidence intervals, are shown in \Cref{fig:angle_vs_corr} as function of the correlation $\correlation$. 

\begin{figure*}
        {
            \newcommand{\quadimgheight}{0.0925\textheight}
            \centering
            \begin{subfigure}[t]{\textwidth}
                \centering
                \includegraphics[height=\quadimgheight,trim=0 10 0 0, clip]{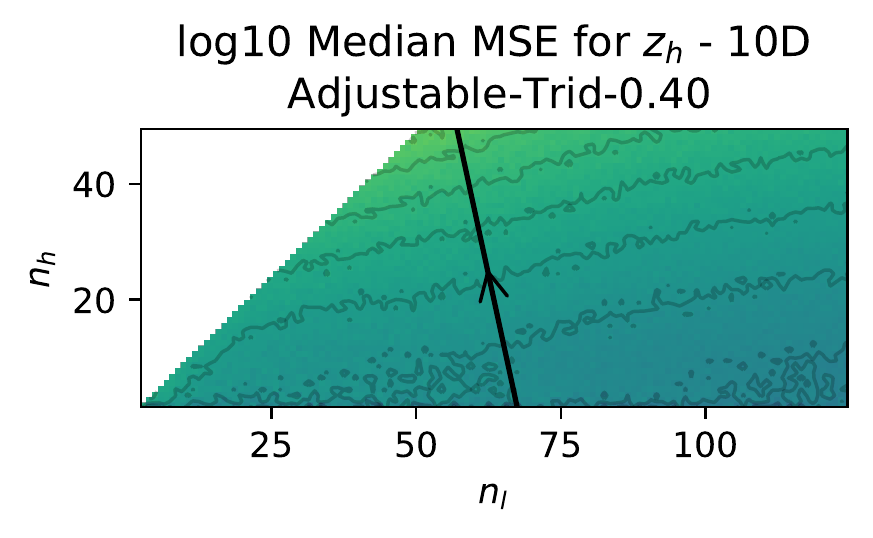} \includegraphics[height=\quadimgheight,trim=0 10 0 0, clip]{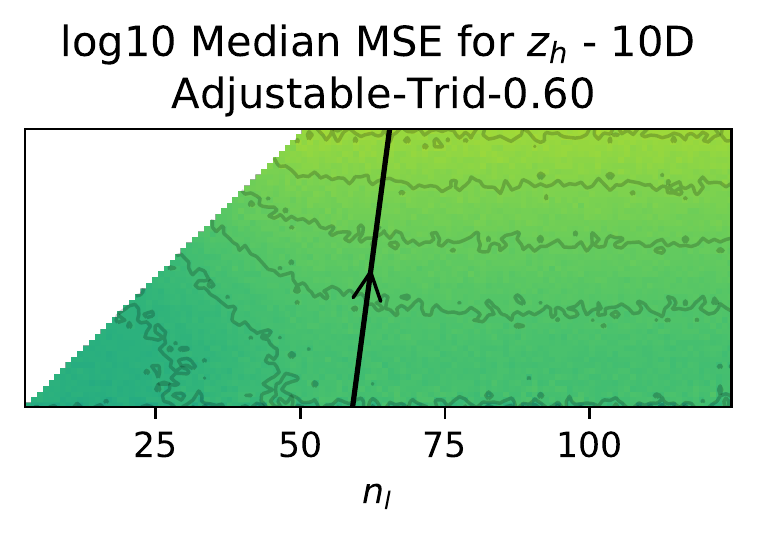}
                \includegraphics[height=\quadimgheight,trim=0 10 0 0, clip]{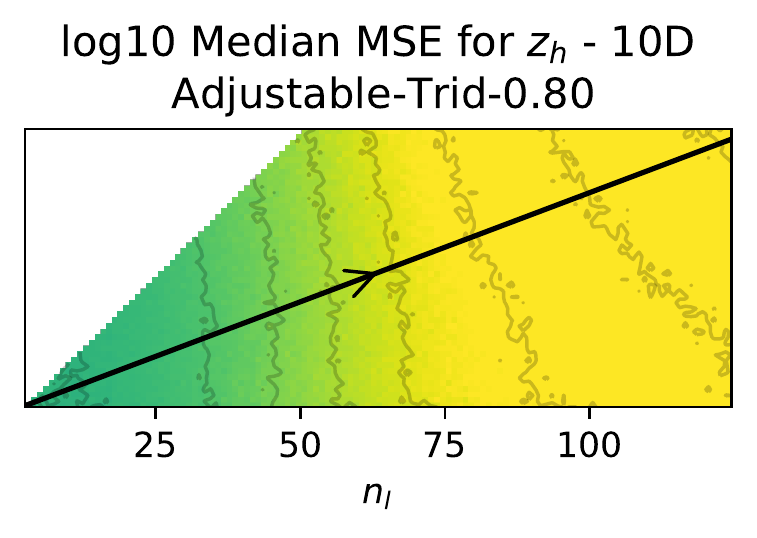}
                \includegraphics[height=\quadimgheight,trim=0 10 0 0, clip]{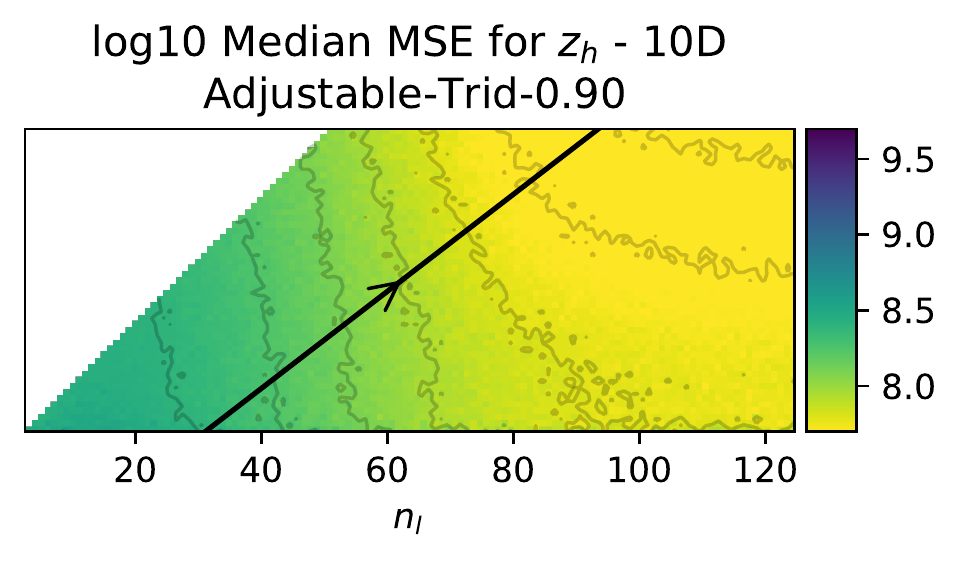}
            \end{subfigure}
            \begin{subfigure}[t]{\textwidth}
                \centering
                \includegraphics[height=\quadimgheight,trim=0 10 0 0, clip]{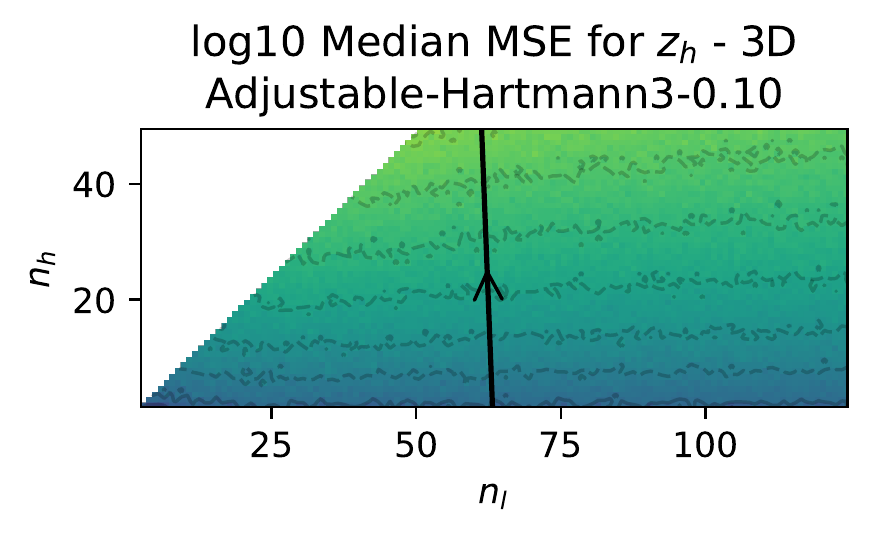}
                \includegraphics[height=\quadimgheight,trim=0 10 0 0, clip]{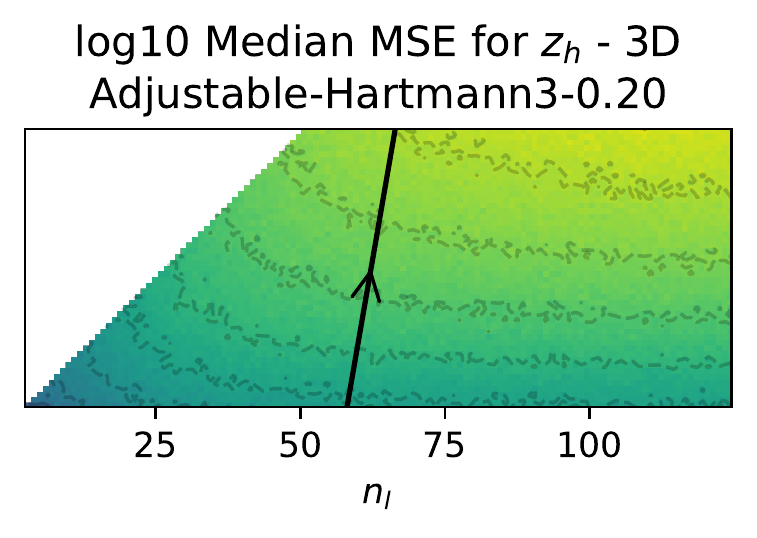} 
                \includegraphics[height=\quadimgheight,trim=0 10 0 0, clip]{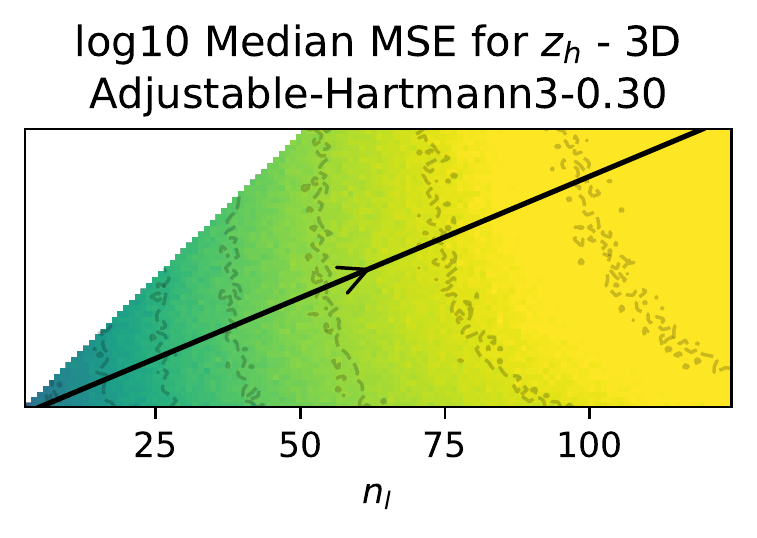} 
                \includegraphics[height=\quadimgheight,trim=0 10 0 0, clip]{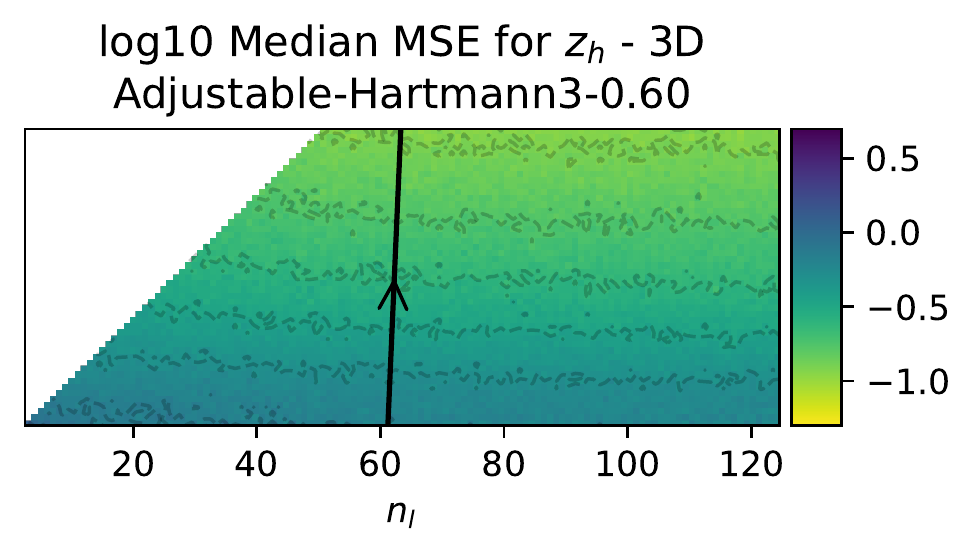} 
            \end{subfigure}
            \vspace{-10pt}
            \caption{
                Error grids for adjustable Trid function for $A=0.4,\;0.6,\;0.8,\;0.9$, with $r=-0.23,\;0.43,\;0.96,\;0.92$, respectively (upper row, from left to right) and the adjustable Hartmann3 function for  $A=0.1,\;0.2,\;0.3,\;0.6$, with $r=0.54,\;0.84,\;0.99,\;0.74$, respectively (lower row, from left to right). The black arrow shows the global gradient direction as described in \Cref{subsec:method-angle-quantification}
                \codelink{2019-10-15-plot-error-grids-adjustables}
                \zoomlink{2019-10-07-adjustables/14061005}
            }
            \label{fig:trid-heatmaps}
        }
        \end{figure*}

        First, it should be noted that the calculated angles cover a very wide range from basically zero up to almost 120 degrees, with the bulk of the values between 30 and 90 degrees. For many functions and in a large range of correlations an angle of around $90\degree$ is computed, which indicates that the accuracy only increases when adding high fidelity samples. Interestingly, angles $\geq90\degree$ are also present. These high angles indicate that the added low-fidelity information actually hurts the accuracy of this hierarchical model, making it perform \emph{worse} than a model trained with fewer low-fidelity samples. This adverse effect of increasing the error by adding low-fidelity samples occurs for correlations up to $r\lessapprox 0.8$ and the tendency gets stronger for less correlated or even anti-correlated benchmark functions. It should be noted that this is not an artifact of the linear fit or the way the error gradient angles are calculated, but truly reflects the behavior of the model error for those functions, as can be seen in \Cref{fig:trid-heatmaps} for the adjustable Trid function.

        It is clear from \Cref{fig:angle_vs_corr} that there is a strong relationship between correlation coefficients and the error gradient angle. For each function the observed gradient angle decreases for higher correlation coefficients. However, the exact values and the functional relationship differ vastly. Even for high correlation coefficients, e.g.,\ $\correlation \geq 0.9$, a large range of resulting error gradient angles can be observed. This means that a high correlation does not necessarily imply a low error gradient angle in the corresponding error grid.

        Furthermore, the Hartmann3 and Trid functions show some other interesting behavior. Note from \Cref{fig:adjustable_correlations} that within $A$'s parameter range (i.e., $[0,1]$), the correlation $r$ for the Hartmann3 and Trid functions is not a bijection. Certain correlation values $\correlation$ are associated with two different gradient angles and vice versa, since different values for $A$ can map to the same correlation $\correlation$ for the Hartmann3 and Trid functions (see \Cref{fig:adjustable_correlations}). The magnified section on the right in \Cref{fig:angle_vs_corr} shows this most clearly.
        To help explain this, recall that the lines in the graph connect data points with adjacent values for $A$, not $\correlation$. This behavior can be visually confirmed by inspecting the error grids directly as shown in \Cref{fig:trid-heatmaps}.

        These examples show that although only a linear fit is used to extract gradient direction, the overall functional dependency of the model error is captured rather well. So, for the purpose of this work, exploring more complicated measures is not necessary. The proposed linear measures are accurate enough to capture the global tendencies, which can already provide insight and the possibility to formulate useful heuristics for practical application (see below).

    

    \subsection{Extrapolation}
    \label{subsec:experiments-extrapolation}
    
        Given that the error grids from subsampling and their subsequent error gradient angles have been shown to match those from full enumeration quite well (see \Cref{app:subsampling}), we propose using this information to answer the question posed in \Cref{sec:problem}: how should additional computation budget be divided between the two available fidelity levels?

        
        As discussed in \Cref{subsec:method-angle-quantification}, we assume the linear fit slope $\gradient$ indicates the direction of most improvement on the error grid. This implies that if the error grid is extended with additional samples, the lowest model errors will be found in the same direction. So, when selecting $\nhigh$ high- and $\nlow$ low-fidelity samples, the ratio between them should match the previously mentioned slope $\gradient$ to achieve the lowest model error, i.e.,


        \begin{equation}
            \label{eq:gradient}
            \frac{\nhigh }{\nlow} = \dgradient
        \end{equation}
        However, in order to apply this method, we need an initial DoE $(\nhighinit, \nlowinit)$ from which to create an error grid and calculate the gradient angle. The number of samples in this initial DoE can be obtained in any way, so does not have to match the ratio $\gradient$. Since we consider the global gradient angle of the error grid when deciding how to select additional samples, we also do not have to select the additional samples $(\Delta\nhigh, \Delta\nlow)$ to bring the total $(\nhigh,\nlow)$ to match the calculated ratio. Instead, we expect the most improvement by having the additional number of samples $(\Delta\nhigh, \Delta\nlow)$ respect the relation
        \begin{equation}
            \label{eq:shifted-gradient}
            \Delta \nhigh = \dgradient \Delta\nlow,  
        \end{equation}
        where $\Delta\nhigh=\nhigh-\nhighinit$ and  $\Delta\nlow=\nlow-  \nlowinit$ are the additional samples to be simulated. A fixed additional budget $b$ can be split between high and low fidelity samples according to the cost ratio $\costratio$,
        \begin{equation}
            \label{eq:budget}
            \Delta\nhigh  +\costratio \Delta \nlow= b.
        \end{equation}
        From \Cref{eq:shifted-gradient,eq:budget} we can determine the best number of additional low and high fidelity samples for a given additional computational budget $b$ as
        \begin{align}
            \label{eq:extrapolation1}
            \Delta\nlow&= \dfrac{b \beta_l }{\beta_h  + \costratio\beta_l},
        \end{align}
        \begin{align}
            \label{eq:extrapolation2}
            \Delta\nhigh&= \dfrac{b \beta_h }{\beta_h  + \costratio\beta_l}.
        \end{align}
   
        \begin{figure}
            \centering
            \includegraphics[trim=0 9 0 5,clip,width=.483\textwidth]{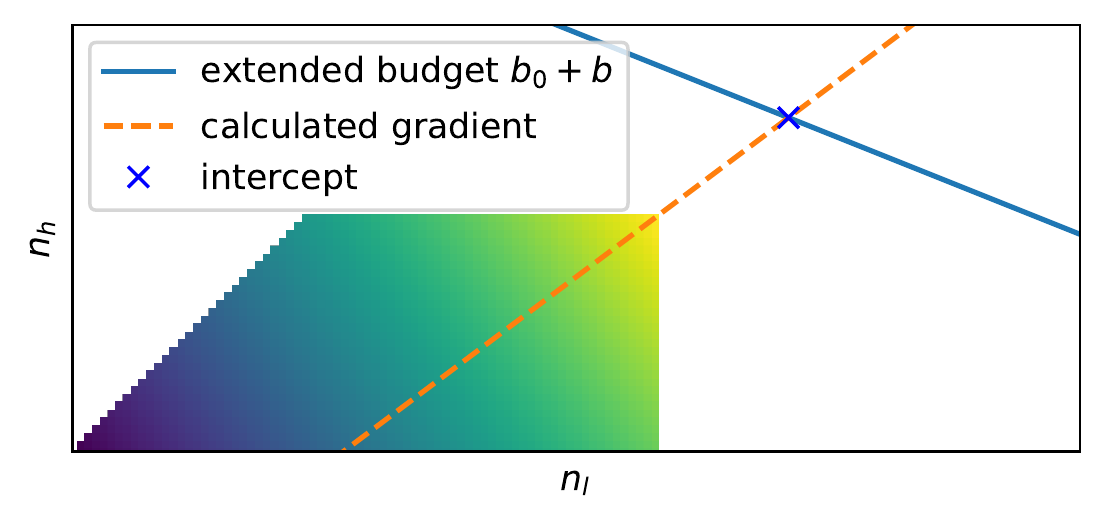}
            \caption{
                Schematic representation of the proposed method to determine the best split for a given additional budget $b$ by extrapolating along the gradient of the error grid, through the upper rightmost point of the error grid until it intersects. Cost ratio for this example is $\costratio=0.4$
                \codelink{2020-07-07-intercept-illustration}
                \zoomlink{2020-07-07-intercept-illustration/14060960}
            }
            \label{fig:extrapolation-intersections}
        \end{figure}

        \Cref{fig:extrapolation-intersections} shows the schematic representation of the proposed extrapolation method for splitting an additional budget $b$. The orange line indicates the extension of the error grid along the direction of the gradient of the error grid. The blue line represents the line where the additional budget is spent according to the cost ratio. The intersection of both lines marks the proposed new samples split for the next DoE. 
        Recall that the initial sample sizes $(\nhighinit, \nlowinit)$ do not need to respect the cost ratio relation of \Cref{eq:budget}, as the initial DoE might be obtained by any method.
        
        To evaluate our method, we consider the example case of starting with a (30, 75) initial DoE that we wish to extend with an additional budget $b=20$, assuming a cost ratio $\costratio=0.4$. First, we create an error grid as described in \Cref{subsec:method-subsampling}, and calculate the gradient as usual. This gradient predicts the division of additional samples according to \Cref{eq:extrapolation1,eq:extrapolation2}, with the size of the resulting DoE falling between (50, 75) for $\Delta\nhigh = 20$ and (30, 125) for $\Delta\nhigh = 0$.
        
\begin{figure*}
            \newcommand{\quadimgheight}{0.0925\textheight}
            \centering

            \includegraphics[width=0.483\textwidth]{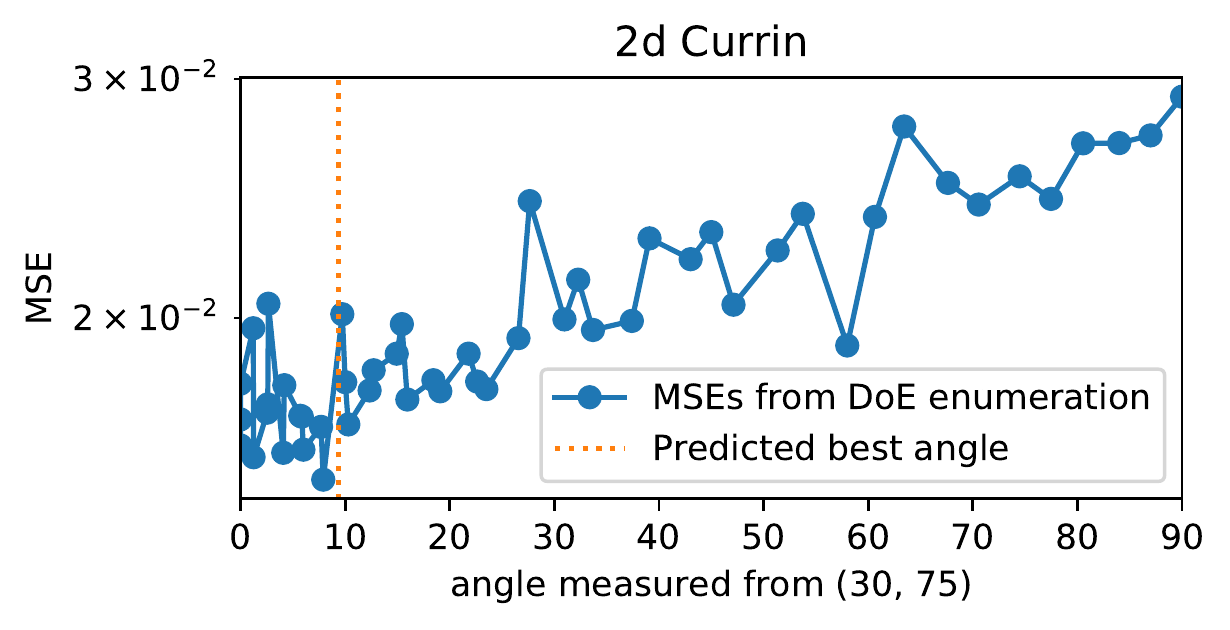}
            \includegraphics[width=0.483\textwidth]{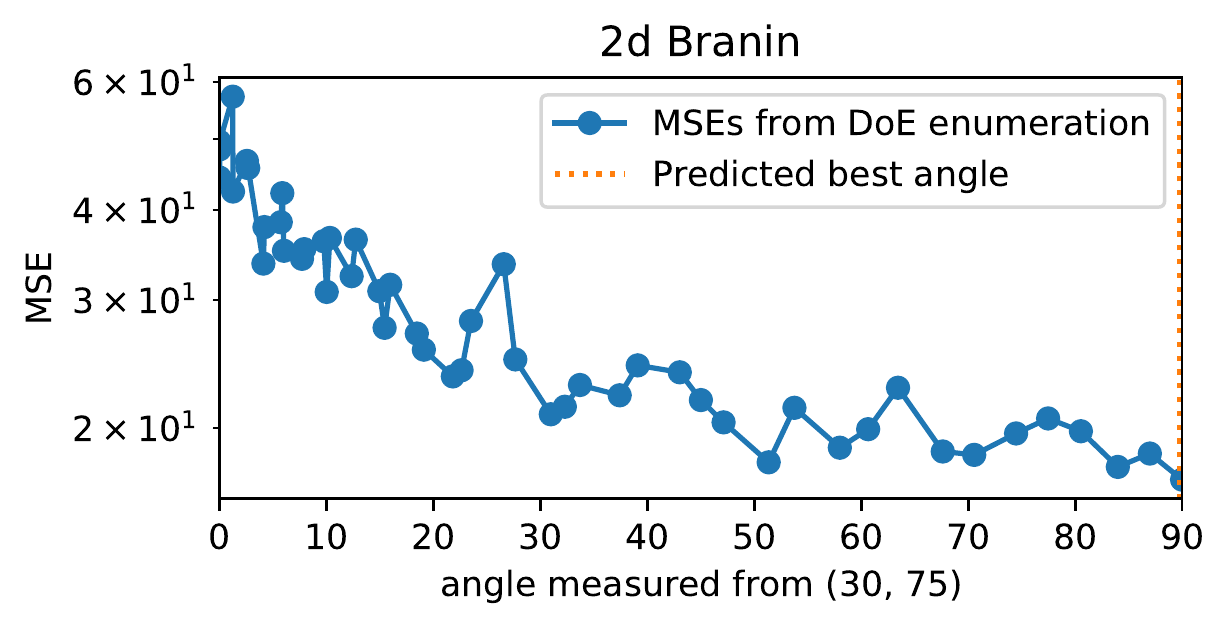}
            \includegraphics[width=0.483\textwidth]{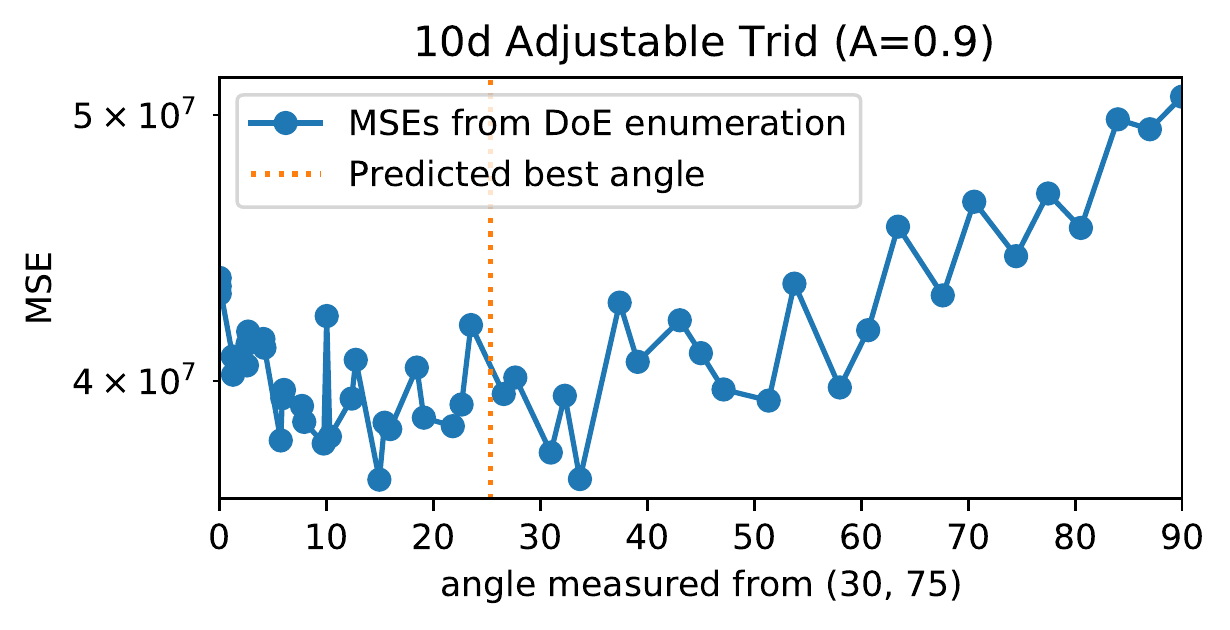}
            \includegraphics[width=0.483\textwidth]{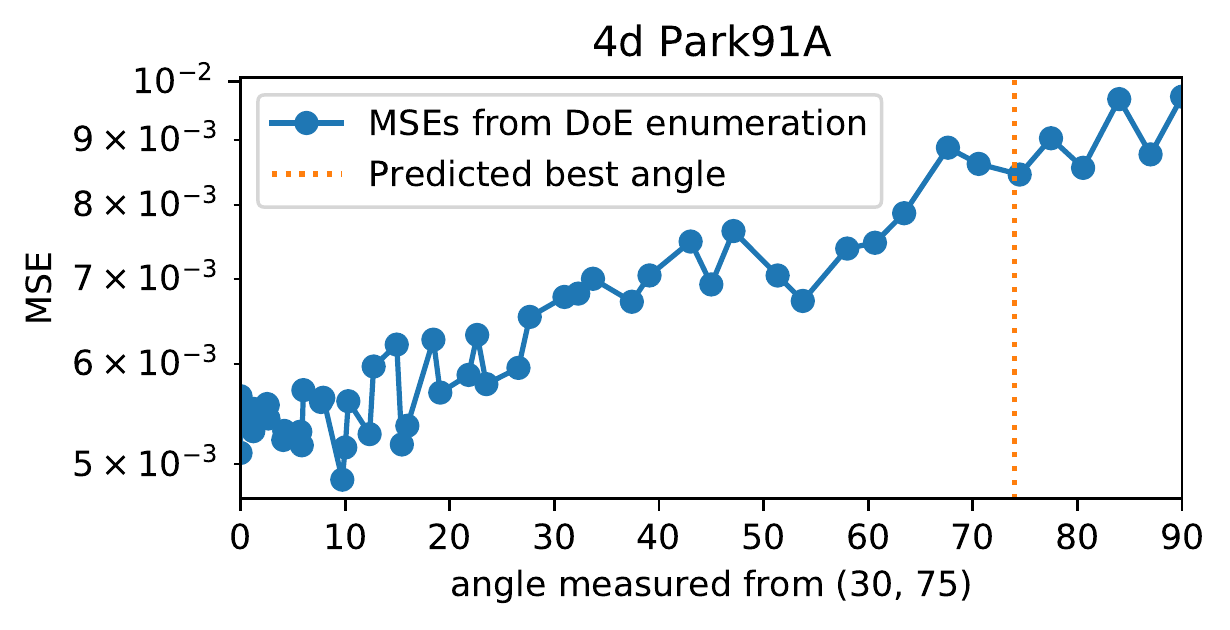}
            \caption{
                Median $\log10$ MSE of DoE sizes along line $\nhigh + \costratio \nlow=80$ $(=b)$ (given $\costratio=0.4$) in the fully enumerated error grid from \Cref{subsec:method-brute-force-enumeration}, shown on the x-axis as the angle measured from the initial sample point $(\nhighinit,\nlowinit)=(30, 75)$ for four benchmark functions. The dashed vertical line shows the angle for the proposed new sample split as calculated by \Cref{eq:extrapolation1,eq:extrapolation2}
                \codelink{2020-07-06-extrapolation}
                \zoomlink{2020-07-06-extrapolation/14061026}
            }
            \label{fig:extrapolations}
        \end{figure*}

        For all DoE sizes between (50, 75) and (30, 125), we can reuse the actual model error data from the full enumeration experiment described in \Cref{subsec:experiments-gradient-angle-analysis}. This data is plotted in \Cref{fig:extrapolations}, as a function of the gradient angle $\theta = \arctan(\frac{\Delta\nhigh}{\Delta\nlow})$, and compared to the predicted gradient angle.

        Generally, it can be observed that the MSE values are quite noisy, even though the median MSE values of 50 different runs are plotted. This is due to the fact that the total number of samples is rather low for the investigated functions, leading to generally large error and also large variations in error. But since we are ultimately interested in real-world problems, where low number of (high fidelity) samples is the norm, we choose such a setup as such noise is expected in these scenarios. The first three examples show that the predicted best angle, i.e., error gradient angle from the subsampled error grid, matches roughly with the minimum measured MSE, regardless of whether this angle is high (90$\degree$) or low (0$\degree$). However, the last plot shows a case where the predicted angle does not lead to a region of low error but rather high error. This is likely an artifact of the linear fit to the error grid. If the error grid has a significantly different gradient in the region of low samples compared to the number of high samples, the linear fit matches the low-sample region and cannot accurately describe the region larger number of samples where the extrapolation is done. So the global estimate of the gradient of the error grid does not align with the local direction of decreasing error around the upper right of the error grid. This is the case for the Park91A function shown in \Cref{fig:example-heatmaps:park91a} and  \Cref{fig:extrapolations}. The error grid of the initial DoE with  $(\nhighinit,\nlowinit)=(30,75)$ is best fit by a linear function with a rather large 75$\degree$\  error gradient angle and consequently the extrapolation suggest to sample $(\Delta\nhighinit,\Delta\nlowinit)=(17,7)$ additional points corresponding to that angle.
        However, only looking at the region above and to the right of the size of the initial DoE with  $(\nhighinit,\nlowinit)=(30,75)$ in the error grid of \Cref{fig:example-heatmaps:park91a} reveals that the direction of decreasing error is more along the $\nlow$ axis., i.e., an angle close to zero.

        This shortcoming of the extrapolation used to determine the split of an additional budget could be mitigated by limiting the region of the error grid to which the fit is done to a smaller area in the upper right of the error grid. By focusing on that region, the strong effect of small sample sizes is excluded, and the linear fit would more accurately model the marginal benefit of adding another sample to the current set.



\section{Conclusions}
\label{sec:conclusion}

    In this work we have empirically examined the trade-off that exists in dividing computational budget between high- and low-fidelity samples in the context of multi-fidelity modeling and optimization problems.

    We presented so-called error grids which are given by the modeling error of a hierarchical surrogate model for a DoE with a given number of high and low fidelity samples $(\nhigh,\nlow)$.  For a complete error grid the modeling error is evaluated for many DoEs with $(\nhigh',\nlow')$ sample points up to the size of the initial DoE, i.e., with $\nhigh'\in[2,\nhigh]$ and $\nlow'\in(\nhigh',\nlow]$.  By this the structure of the model error is revealed and the behavior of the modeling error as function of the split between high and low fidelity samples can be analyzed.

    We captured the global trend in the modeling error by  fitting a linear hyperplane through the $\log_{10}$ of the Mean Squared Errors. The linear fit easily lends itself to extracting the error gradient's global direction, which is used to identify the global direction of reducing error in the $\nhigh$-$\nlow$ plane. We analyzed the error grids for a multitude of benchmark functions, where some functions have a parameter which allows the tuning of the relation between low and high fidelity functions.

    The first version of the error grid we presented uses an independently sampled DoE for each hierarchical model with a given sample split. This requires a very large number of independent function evaluations and we therefore presented a simple subsampling method which needs only the available evaluations of an initial DoE in the spirit of cross-validation.  We showed for multiple benchmark functions that the direction of the gradient of the error grid can be estimated from the subsampling error grid reasonably well.

    Based on the extracted global direction of the gradient of the modeling error, we proposed a simple scheme which allows an informed decision about how to divide additionally available evaluation budget between the different fidelities. We showed that the scheme works well on most benchmark functions. Those cases where the predicted split of the additional budget did not extrapolate to a region with smaller model error are characterized by a change of the dominant behavior of the error grid with the number of high- and low-fidelity samples. This shortcoming of the proposed method could be mitigated by performing the linear fit only to the region with the highest numbers of samples (i.e.\ the upper right part of the error grid) to increase the influence of the region of interest and at the same time reduce the sensitivity to low number of samples.

    We see two main applications: First, as a means to characterize the behaviour of a fidelity level with respect to its accuracy.   
    Rather than relying on heuristics based on the correlation between the fidelity levels, we propose that error grids can provide valuable initial insight into the benefit of additional samples from each fidelity level. Secondly, as a possible means of online fidelity selection for multi-fidelity optimization use cases. The proposed approach determines the optimal division between number of high- and low-fidelity samples given a set of samples. This can be utilized at each iteration of an optimization procedure to determine the split between high and low fidelity of the newly generated samples. The  marginal benefit of each fidelity level will be reflected in the error grid's gradient direction and thereby steering the fidelity selection for the optimization.

    In future work, experiments on additional benchmark functions and also real-world problems will have to be performed in order to confirm the benefits of the error grids for those applications. As we have only considered a hierarchical surrogate model based on a simplified additive co-kriging design, other multi-fidelity models should be investigated, since we expect the gradient angle to change according to the quality of the model fit. Additionally, the benefit of using the error grids and the extrapolation scheme for optimization use cases needs to be explored. 


\section*{Conflict of Interest}
    On behalf of all authors, the corresponding author states that there is no conflict of interest.

\section*{Acknowledgements}
    This work is part of research program DAMIOSO, project number 628.006.002, which is partly financed by the Netherlands Organisation for Scientific Research (NWO).

\section*{Data Availability Statement}
    Source code \cite{mlcs_code_zenodo} and data files \cite{mlcs_data_zenodo} are publicly archived on Zenodo.


%


\bibliographystyle{tfcad}
\bibliography{references}

\sectionbreak

\appendix

\section{Gradient Angle Confidence Interval}
\label{app:angle-ci}

    From the linear fit of equation \eqref{eq:lin-reg} we can also calculate the standard errors for the linear fit parameters $\beta_i$ associated with input feature $n_i$, i.e., the number of high- or low-fidelity samples $\nhigh$ or $\nlow$, as
    \begin{equation}
        \label{eq:standard-error}
        se_{\beta_i} = 
        \dfrac{\sqrt{\dfrac{\text{SSE}}{N_\mathrm{DoE}-df}}}{\sqrt{\sum (n_i - \overline{n_i})^2}} = 
        \dfrac{\sqrt{\dfrac{\sum (f_h(x) - z_h(x))^2}{N_\mathrm{DoE}-df}}}{\sqrt{\sum (n_i - \overline{n_i})^2}},
    \end{equation}
    where $n_i$ can either be the number of high- or low-fidelity samples, $\overline{n_i}$ is the respective mean, $df$ is the number of degrees of freedom, i.e., the number of samples $N_\mathrm{DoE}$ minus three for the number of parameters from the linear regression equation ($\alpha, \beta_h, \beta_l$), and SSE is the sum of squared errors for our linear fit.

    Using these standard errors we can determine a 95\% Confidence Interval (CI) for the slope $\frac{\beta_h}{\beta_l}$, from which we can estimate the range of the angle:

    \begin{equation}
        \label{eq:slope-ci}
        \text{CI\ } \frac{\beta_h}{\beta_l} = \frac{\beta_h}{\beta_l} \pm 1.96 \sqrt{\left(\frac{se_{\beta_h}}{\beta_h}\right)^2 + \left(\frac{se_{\beta_l}}{\beta_l}\right)^2}
    \end{equation}
    \begin{align}
        \label{eq:angle-ci}
        \nonumber
        &\text{CI\ } \theta
        \approx \Bigg[ \tan^{-1} \left(\frac{\beta_h}{\beta_l} - 1.96 \sqrt{\left(\frac{se_{\beta_h}}{\beta_h}\right)^2 + \left(\frac{se_{\beta_l}}{\beta_l}\right)^2}\right),\\
        &\tan^{-1}\left(\frac{\beta_h}{\beta_l} + 1.96 \sqrt{\left(\frac{se_{\beta_h}}{\beta_h}\right)^2 + \left(\frac{se_{\beta_l}}{\beta_l}\right)^2}\right)\Bigg]
    \end{align}
     
    Using this CI, we can be more certain of the global error gradient angle of the error grid. In any local section of the error grid, the angle can still be significantly different, but the proposed method provides a robust estimate of the average gradient and serves the purpose of discriminating the global behavior of different benchmark functions.

\section{Subsample Analysis}
\label{app:subsampling}

    To validate our proposed method of reducing the number of necessary function evaluations for the error grid analysis described in \Cref{subsec:method-subsampling}, we explore the influence of the sizes of training and test sets.
    Results of the following three setups are compared:
    \begin{enumerate}
        \item Independent full enumeration of training and test sets as described in \Cref{subsec:method-brute-force-enumeration};
        \item Subsampled training set and independent test set;
        \item Subsampled training set and left-over test set, i.e., full subsampling as described in \Cref{subsec:method-subsampling}
    \end{enumerate}
    If accurate enough, the third setup is the preferred approach for practical applications, as it uses any computational budget most efficiently by using each available evaluation for either test or training sets, and no further evaluations besides that.

    Comparing between 1. and 2. shows the dependence of the procedure's results on the initial DoEs used as training set: does the spread of the subsamples cover the search space well enough to simulate independent DoEs of the subsample size?     
    The comparison between 2. and 3. illustrates how accuracy tests with (much) less information influence the results.

    \begin{figure*}
        \centering
        \begin{subfigure}[t]{\textwidth}
            \includegraphics[width=\textwidth,trim=0 10 0 0, clip]{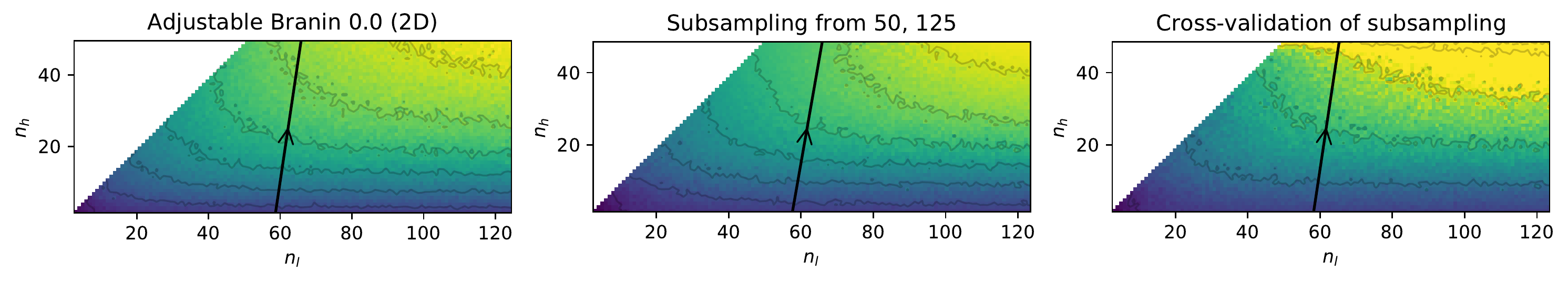}
        \end{subfigure}
        \begin{subfigure}[t]{\textwidth}
            \includegraphics[width=\textwidth,trim=0 10 0 0, clip]{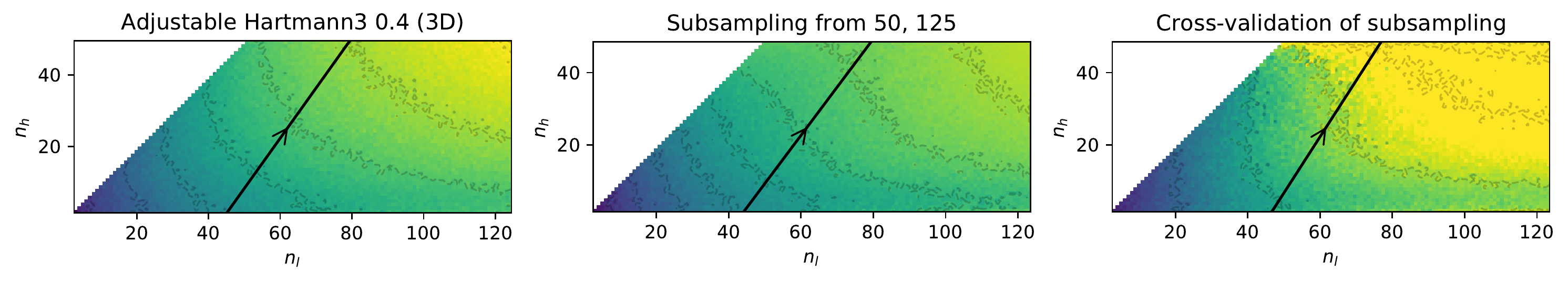} 
        \end{subfigure}
        \begin{subfigure}[t]{\textwidth}
            \includegraphics[width=\textwidth,trim=0 10 0 0, clip]{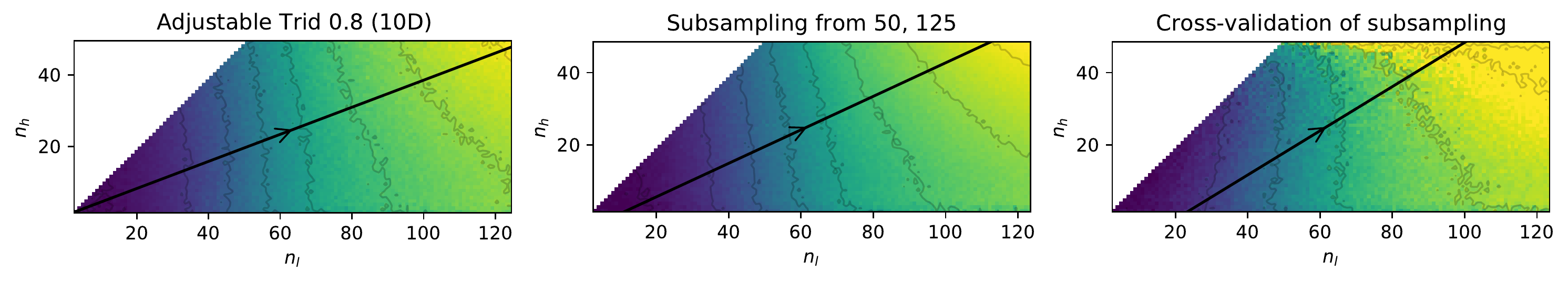} 
        \end{subfigure}
        \caption{
            \textbf{Comparison of error grids} for the Adjustable Branin ($A = 0.0$, $r=0.99$), Hartmann3 ($A = 0.4$, $r=0.97$), and Trid ($A = 0.8$, $r=0.96$) using different methods.  \textbf{Left}: error results using independent training and test set (\Cref{subsec:method-brute-force-enumeration}). The estimated error gradient angles, as illustrated by the black arrows, are $\theta = 81.8\degree$, $55.4\degree$, $20.4\degree$. \textbf{Middle}: error results using subsampled training set, with independent test set.  The estimated error gradient angles are $\theta =  80.3\degree$, $54.0\degree$, $25.5\degree$. \textbf{Right}: error results using subsampled training set and left-over test set (\Cref{subsec:method-subsampling}).  The estimated error gradient angles are $\theta = 81.7\degree$, $58.0\degree$, $32.2\degree$.
            \codelink{2020-03-09-adjustables-subsampling-comparisons}
            \zoomlink{2020-03-09-adjustables-subsampling-comparisons/14061020}
        }
        \label{fig:tri-compare-subsampling}
    \end{figure*}  

    \Cref{fig:tri-compare-subsampling} shows example comparisons for the adjustable Branin, Hartmann3 and Trid functions. The ground-truth error grid (left) shows a mostly $90\degree$ gradient for $\nlow\gg \nhigh$ with a trend toward 45$\degree$~near the $\nhigh = \nlow$ diagonal.
    The subsampling error grids (middle and right panels) are visibly noisier than the ground-truth, but show similar shape characteristics. Despite the subtle differences in curvature, the resulting error gradient angles are very similar between 80$\degree$-82$\degree$.

    \begin{table}
        \centering
        \begin{tabular}{@{}ll@{}}
    \toprule
    Function & Parameter values used for $A_1, \dots, A_4$\\
    \midrule
    Branin    & 0.00, 0.05, 0.25 \\
    Paciorek  & 0.05, 0.10, 0.15, 0.20, 0.25 \\
    Hartmann3 & 0.20, 0.25, 0.30, 0.35, 0.40 \\
    Trid      & 0.65, 0.70, 0.75, 0.80, 0.85, 0.90, 0.95, 1.00 \\
    \bottomrule
\end{tabular}
        \caption{
            Listing of all parameters that are used to compare correlation between error gradient angles from full enumeration and subsampling
        }
        \label{tab:corr_comparison_parameters}
    \end{table}  
    
    \begin{figure}
        \centering
        \begin{subfigure}[t]{.33\textwidth}
            \includegraphics[width=\textwidth]{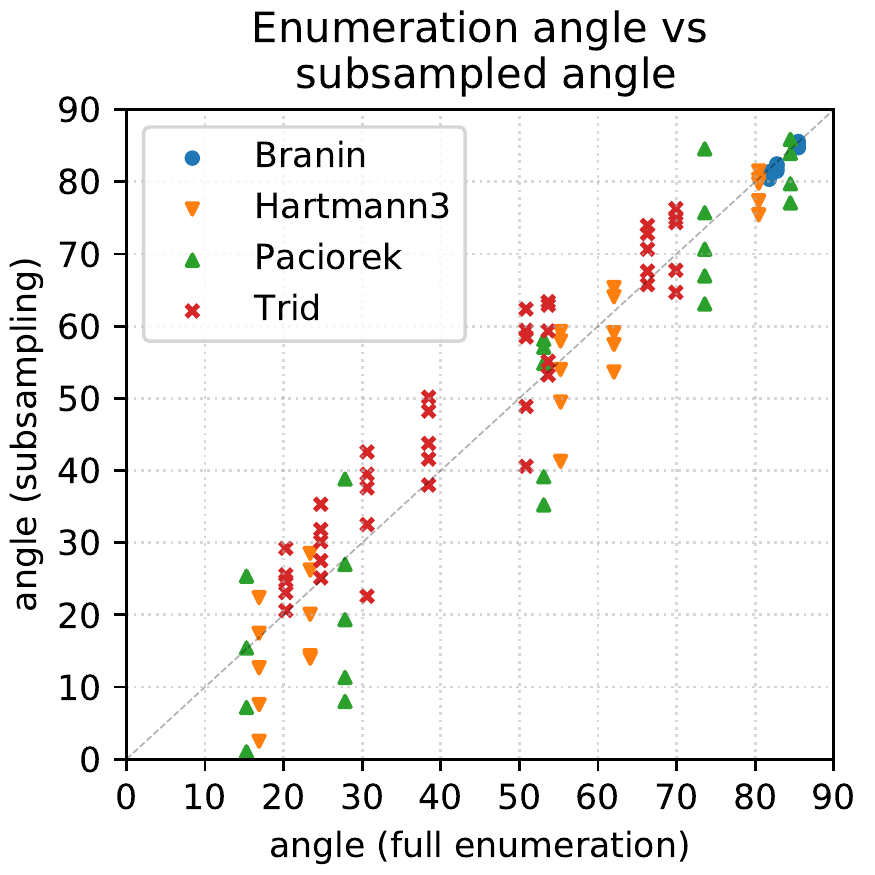}
            \caption{
                Subsampled training set, external test set
            }
            \label{subfig:scatter-compare-partial-subsampling}
        \end{subfigure}
        \begin{subfigure}[t]{.02\textwidth}
        \end{subfigure}
        \begin{subfigure}[t]{.33\textwidth}
            \includegraphics[width=\textwidth]{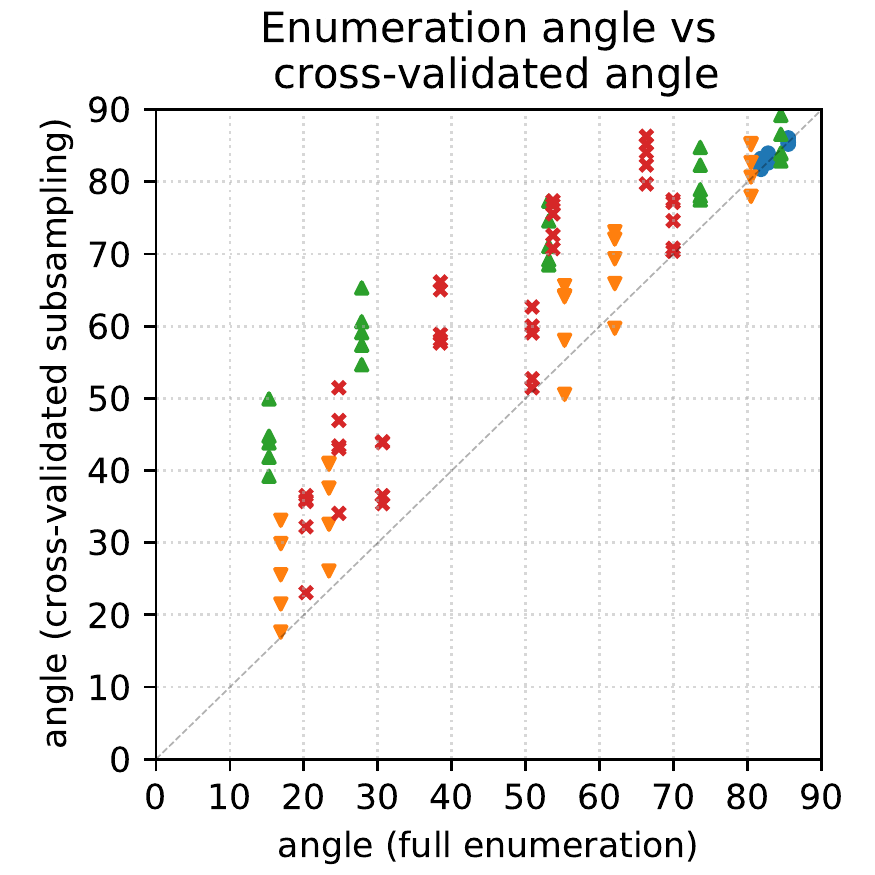}
            \caption{
                Subsampled training set, left-over test set
            }
            \label{subfig:scatter-compare-full-subsampling}
        \end{subfigure}
        \caption{
            \textbf{Error gradient angle correlation} The horizontal axis shows the angles as determined using the full enumeration procedure, while the vertical axis shows the angles calculated by the procedure mentioned by the caption below each figure
            \codelink{2020-03-09-subsampling-gradients}
            \zoomlink{2020-03-09-subsampling-gradients/14061023}
        }
        \label{fig:scatter-compare}
    \end{figure}  

    For a more detailed analysis, we select 21 (function, parameter) combinations as described in \Cref{tab:corr_comparison_parameters} that cover a wide range of error gradient angles. For all cases, both subsampling procedures are repeated five times using independent initial DoEs in each case, and the error gradient angles are calculated. \Cref{fig:scatter-compare} shows the correlation between the angles from the full enumeration and the subsampling procedures, as evaluated using both the independent large test set (left) and with the full subsampling approach with left-over test set (right).

    First considering \Cref{subfig:scatter-compare-partial-subsampling}, we see that the angles from subsampling correspond very well to the ground-truth angles, with a spread of $\pm 5\degree$ to $\pm 15\degree$ roughly symmetrically around diagonal. The magnitude of the spread is visibly larger as the angles are smaller.

    \Cref{subfig:scatter-compare-full-subsampling} compares the ground-truth error gradient angles with those calculated using the test set based on cross-validation. Again, the variance in the estimated angle becomes smaller for larger angles.
    It is interesting to note that the spreads of the angles are also roughly between $\pm 5\degree$ to $\pm 15\degree$, but there  seems to be a systematic shifted of the angles estimated by full subsampling with a cross-validated test set to \emph{higher} values. The exact shift is dependent on the underlying function, e.g.,\ the Paciorek function having a much larger shift than the Hartmann3 function. So while not an exact prediction of the ground-truth error gradient angle, it can be interpreted as a worst-case estimate: the ground-truth error gradient angle is unlikely to be higher than what results from this procedure.


\end{document}